\documentclass[sn-nature]{sn-jnl}% Math and Physical Sciences Numbered Reference Style 

\usepackage{graphicx}%
\usepackage{multirow}%
\usepackage{amsmath,amssymb,amsfonts}%
\usepackage{amsthm}%
\usepackage{mathrsfs}%
\usepackage[title]{appendix}%
\usepackage{xcolor}%
\usepackage{textcomp}%
\usepackage{manyfoot}%
\usepackage{booktabs}%
\usepackage{algorithm}%
\usepackage{algorithmicx}%
\usepackage{algpseudocode}%
\usepackage{listings}%
\graphicspath{{figures/}}

\theoremstyle{thmstyleone}%
\theoremstyle{thmstyletwo}%

\theoremstyle{thmstylethree}%

\begin{document}

\title[Article Title]{Accurate Mediterranean Sea forecasting via graph-based deep learning}


\author*[1,2]{\fnm{Daniel} \sur{Holmberg}}\email{daniel.holmberg@helsinki.fi}
\author[2]{\fnm{Emanuela} \sur{Clementi}}\email{emanuela.clementi@cmcc.it}
\author[2,3]{\fnm{Italo} \sur{Epicoco}}\email{italo.epicoco@cmcc.it}
\author[1]{\fnm{Teemu} \sur{Roos}}\email{teemu.roos@helsinki.fi}

\affil[1]{\orgdiv{Department of Computer Science}, \orgname{University of Helsinki}, \country{Finland}}
%\affil[2]{\orgdiv{Regional Ocean Forecasting Systems Division}, \orgname{Euro-Mediterranean Center on Climate Change}, \country{Italy}}
\affil[2]{\orgname{CMCC Foundation - Euro-Mediterranean Center on Climate Change}, \country{Italy}}
\affil[3]{\orgdiv{Department of Engineering for Innovation}, \orgname{University of Salento}, \country{Italy}}

% \affil[3]{\orgdiv{Department}, \orgname{Organization}, \orgaddress{\street{Street}, \city{City}, \postcode{610101}, \state{State}, \country{Country}}}

\abstract{Accurate ocean forecasting systems are essential for understanding marine dynamics, which play a crucial role in sectors such as shipping, aquaculture, environmental monitoring, and coastal risk management. Traditional numerical solvers, while effective, are computationally expensive and time-consuming. Recent advancements in machine learning have revolutionized weather forecasting, offering fast and energy-efficient alternatives. Building on these advancements, we introduce SeaCast, a neural network designed for high-resolution regional ocean forecasting. SeaCast employs a graph-based framework to effectively handle the complex geometry of ocean grids and integrates external forcing data tailored to the regional ocean context. Our approach is validated through experiments at a high horizontal resolution using the operational numerical forecasting system of the Mediterranean Sea, along with both numerical and data-driven atmospheric forcings. Results demonstrate that SeaCast consistently outperforms the operational model in forecast skill, marking a significant advancement in regional ocean prediction.}

\keywords{Regional ocean forecasting, graph neural networks, learned simulation}

\maketitle

\section{Introduction}
\label{sec:introduction}

Predicting sea dynamics is a formidable scientific challenge, driven by the need to anticipate changes in ocean conditions that affect weather systems, marine ecosystems, and a wide range of maritime activities~\cite{le2019cmems}. While the need for improved ocean and coastal data is global, actionable decisions in sectors such as shipping, marine resource management, and coastal planning often depend on regional, high-resolution models that can deliver accurate forecasts tailored to local conditions~\cite{sun2019skrips, sakamoto2019development, ciliberti2021blacksea, karna2021nemonordic, zhu2022scsofs}.

\begin{figure*}[ht]
\centering
\includegraphics[width=\textwidth]{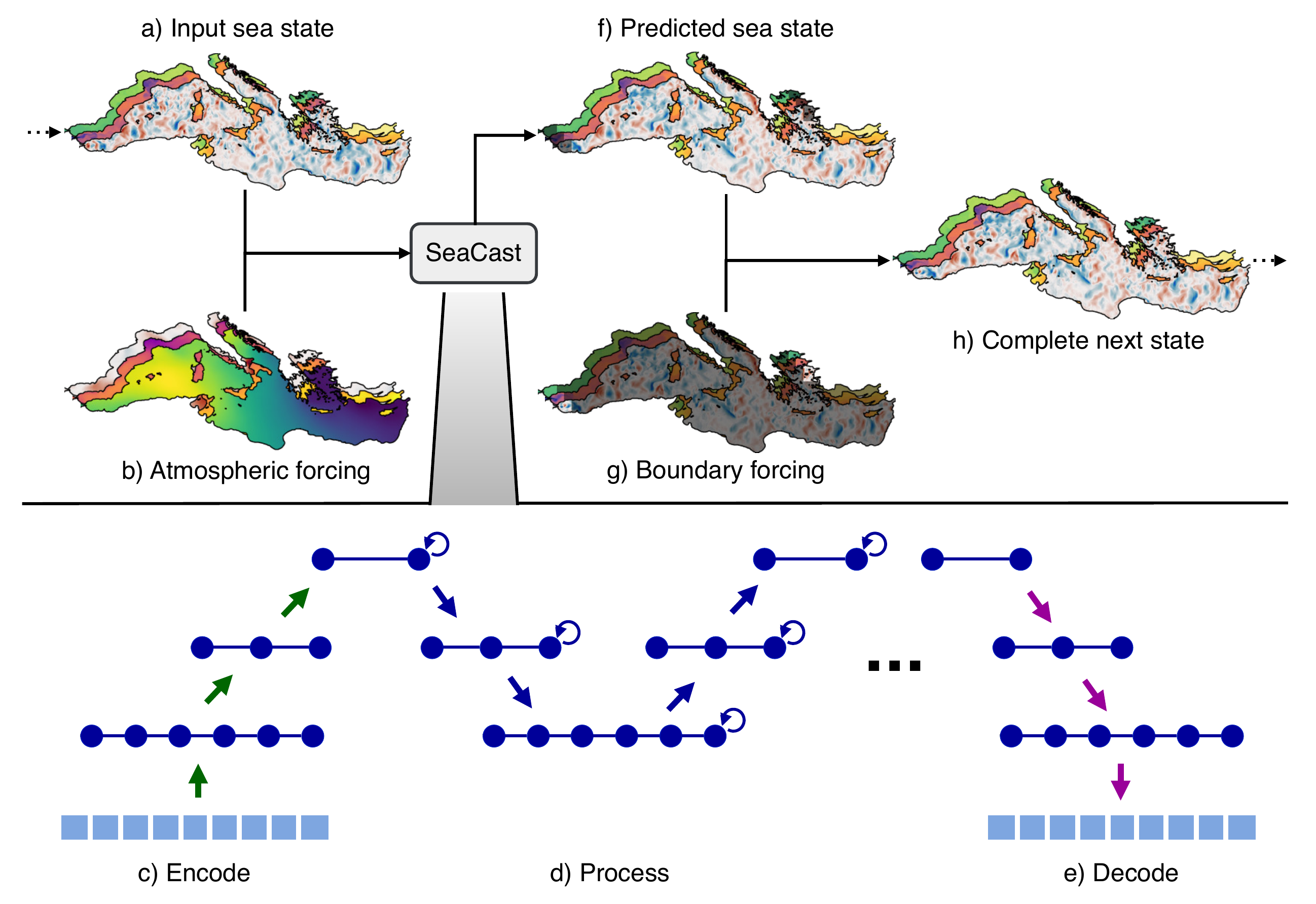}
\caption{SeaCast performs autoregressive ocean forecasting using a graph neural network (GNN). (a) The input sea state and (b) atmospheric forcing are projected onto a coarser mesh representation through the (c) encoder. (d) GNN layers process this latent representation on a hierarchical mesh. The output is then (e) decoded back onto the original grid to form (f) a predicted sea state. (g) Boundary conditions (here exaggerated in size) are incorporated to produce (h) the complete next state. This forecast is then fed back into the system as a new input (as shown by the arrow looping from h to a), enabling multi-step forecasting through repeated application of the encode–process–decode cycle.}
\label{fig:schema}
\end{figure*}

Here, we focus on the Mediterranean Sea, a region characterized by intense mesoscale activity, complex coastlines, and dense coastal populations. Regional ocean forecasts in this area are important for managing marine operations and mitigating coastal risks. Accurate and timely predictions empower local agencies to address a wide range of challenges, from water quality monitoring~\cite{bruschi2021indexes} and oil spill containment~\cite{liubartseva2023oilspill} to maritime route optimization~\cite{mannarini2024visir}. Currently, the \textit{Mediterranean Forecasting System} (MedFS)~\cite{coppini2023forecast} delivered through the Copernicus Marine Service (CMEMS) provides medium-range (10 days) forecasts using a physics-based two-way coupled wave-current modeling system, and serves as the operational reference system for the region.

Despite their accuracy, numerical models such as MedFS are computationally intensive and time-consuming to run. Recent advancements in machine learning-based weather prediction (MLWP) have demonstrated the potential for fast and performant alternatives. Several autoregressive machine learning models now rival or surpass state-of-the-art physics-based systems used by meteorological agencies worldwide. Notable developments include architectures based on Transformers~\cite{bi2023panguweather, nguyen2024scaling}, neural operators~\cite{pathak2022fourcastnet}, and graph neural networks (GNNs)~\cite{keisler2022forecasting, lam2023graphcast, oskarsson2024probabilistic}.

While these advances have been transformative in the atmospheric domain, its application to ocean forecasting remains relatively nascent. Existing work has focused mostly on forecasting the global ocean, at climate~\cite{dheeshjith2025samudra}, seasonal~\cite{wang2024coupled, guo2024data}, and medium-range~\cite{wang2024xihe, aouni2024glonet, cui2025forecasting} timescales. Regional studies to date have primarily targeted the forecasting of a small set of surface variables, such as sea ice~\cite{andersson2021seasonal}, sea surface height (SSH)~\cite{chattopadhyay2024oceannet}, or surface currents and temperature~\cite{subel2024building}. However, regional data-driven ocean forecasting systems for depth-resolved prediction in high resolution operational settings are still lacking.

To address this gap for the Mediterranean Sea, we introduce \emph{SeaCast}, an autoregressive machine learning model designed for regional ocean forecasting, extending the methodology from weather forecasting using hierarchical GNNs~\cite{oskarsson2024probabilistic}. Our approach involves several key features that enable accurate prediction of ocean states: 1) we adapt the graph creation, training, and evaluation processes to accommodate the irregular geometry of ocean grids; 2) the model includes relevant atmospheric forcing near the sea surface; and 3) lateral boundary forcing is applied to account for water inflow and outflow, ensuring compatibility with the ocean at large. SeaCast is evaluated in an operational setting alongside MedFS, including validation against observational data. In addition, we conduct targeted experiments to assess the influence of atmospheric forcing components and the effects of training period length.

\section{Results}
\label{sec:results}

\subsection{SeaCast: data-driven regional ocean forecasting}

SeaCast is a data-driven forecasting model for the Mediterranean Sea that delivers 15-day forecasts across 18 depth levels on a $1/24^\circ$ (approximately 4\,km) horizontal grid, matching the resolution of the operational MedFS system (see Supplementary~A). The model predicts key physical ocean variables, including depth-resolved zonal and meridional currents, salinity, and temperature, along with SSH, resulting in 73 predicted fields in total. A major advantage of SeaCast is its computational efficiency: it produces a full 15-day forecast in just 20 seconds on a single GPU, compared to approximately 70 minutes required by MedFS to generate a 10-day forecast on 89 CPU cores, using a 120-second timestep and producing outputs at 141 depth levels. Although the two systems differ in how they operate, the data-driven approach represents a significant speedup in producing upper ocean forecasts compared to what was previously possible.

The model architecture follows an encode–process–decode framework~\cite{sanchez2020learning} operating on a hierarchical graph mesh tailored to the Mediterranean basin (see Supplementary~B). As illustrated in Figure~\ref{fig:schema}, the input sea state and atmospheric forcing are first encoded onto a coarser multi-resolution mesh representation. The latent features are then processed by GNN layers in a hierarchical fashion, enabling the model to capture both short- and long-range ocean interactions. The processed output is subsequently decoded back onto the original high-resolution grid. Rather than directly predicting the next sea state, the model learns the tendency, i.e. the expected change over a one-day interval, which is added to the current state to obtain the forecast. Dynamic boundary conditions are then incorporated to generate the complete next sea state. This predicted state is then fed back into the model as input for the next step, enabling the model to produce forecasts at different lead times by repeating the application of the same cycle in an autoregressive manner. Unlike multiscale models such as GraphCast~\cite{lam2023graphcast}, which connect nodes on a single mesh level, the hierarchical approach we employ separates the domain into multiple, distinct mesh levels. This design results in more uniform connectivity from the mesh to the grid, helping to mitigate artifacts associated with nodes having variable neighborhood sizes~\cite{oskarsson2024probabilistic}.

The atmospheric forcing used in SeaCast includes 10-meter wind stress components, 2-meter temperature, and mean sea level pressure (MSLP), combined with the sine and cosine of the day of year as seasonal indicators. Open boundary conditions are handled by overwriting the predicted sea state with the true state in boundary regions located at the Strait of Gibraltar and the Dardanelles Strait during training, and with MedFS data during evaluation. This approach is necessary to ensure realistic representation of inflow and outflow dynamics for the region considered~\cite{marchesiello2001open}.

SeaCast is trained on 35 years (1987–2021) of Mediterranean reanalysis daily mean data~\cite{escudier2021reanalysis} and fine-tuned using two additional years (2022–2023) of daily operational analysis~\cite{coppini2023forecast}. Fine-tuning serves multiple purposes: it exposes the model to more recent sea states, enables the use of analysis fields as initial conditions in an operational setting, and allows the model to adapt to updates introduced in the operational MedFS system that are absent in the reanalysis. The atmospheric forcings driving the surface of the model are taken from the European Centre for Medium-Range Weather Forecasts (ECMWF) ERA5 reanalysis data~\cite{hersbach2023era5} during the training phase, and during testing either from the ECMWF \emph{ensemble control forecast} (ENS)~\cite{molteni1996ens} or the recent \emph{artificial intelligence forecasting system} (AIFS)~\cite{lang2024aifs}. SeaCast is evaluated on a daily test set with initializations spanning from the beginning of July 2024 to the end of December 2024, each producing a 15-day forecast. Performance is assessed using both model-based reference fields and satellite observations, and benchmarked against the operational MedFS. SeaCast uses the same initial conditions available to MedFS making for a fair comparison. Further details on the method are provided in Section~\ref{sec:methods}.

\subsection{Comparing SeaCast to the operational MedFS}

SeaCast is evaluated against the operational MedFS, the primary regional ocean forecasting system for the Mediterranean Sea. MedFS provides deterministic forecasts up to 10-days lead, adhering to the 10-day standard of CMEMS ocean products, and also the extent of ECMWF's legacy high-resolution (HRES) atmospheric forecasts used as surface forcing. Recent changes to the ECMWF Integrated Forecasting System (IFS), extending medium-range predictions from 10 to 15 days in Cycle 48r1~\cite{ecmwf2024plans}, was taken into account when developing SeaCast. By default, the model uses AIFS as atmospheric forcing, and in Section~\ref{sec:atm_forcing}, also ENS for comparison. By using these products, and persisting the lateral boundary conditions the last 5 days, SeaCast is capable of producing 15-day forecasts. This extended forecast horizon represents an important step forward, enabling potential earlier warnings of marine extremes and enhancing medium-range planning capabilities.

To benchmark performance, we evaluate the forecast skill of both SeaCast and MedFS across six key ocean variables: zonal and meridional currents, salinity, temperature, sea surface temperature (SST), and sea level anomaly (SLA). Subsurface variables are validated against daily mean MedFS analysis fields. These gridded analysis fields, which assimilate quality-controlled observations into the physics model, provide a stable reference for subsurface conditions. SST and SLA are compared to Level-3 (L3) satellite observations. Specifically, SST predictions from the model's uppermost layer are evaluated against merged multi-sensor satellite data at $1/16^\circ$ resolution~\cite{nardelli2013sstobs}, while SSH forecasts are converted to SLA and compared to along-track 5\,Hz altimeter measurements from multiple satellite missions (see Section~\ref{sec:satellite_data} for further details).

\begin{figure*}[h]
\centering
\includegraphics[width=\textwidth]{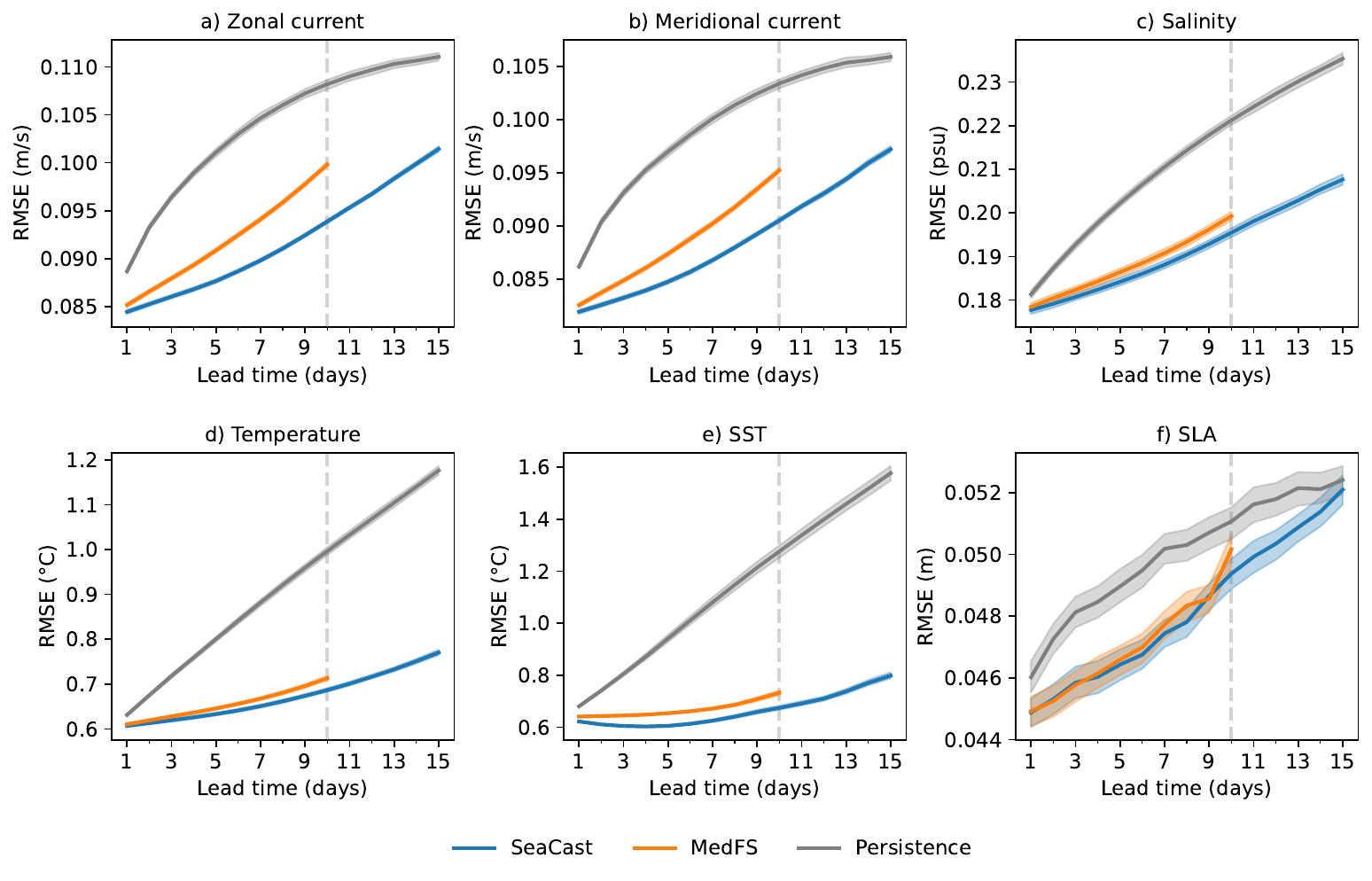}
\caption{RMSE as a function of lead time for SeaCast, MedFS, and a persistence baseline. Sea state variables with several vertical levels (zonal and meridional currents, salinity, and temperature) are evaluated against analysis fields where the error is an average over all depths. SST and SLA are validated against L3 satellite observations, using the uppermost modeled temperature level for the SST comparison. The shading corresponds to the $50\%$ confidence intervals estimated via bootstrapping.}
\label{fig:rmse}
\end{figure*}

A persistence baseline is included as a naive benchmark, generated by repeating the initial conditions over all lead times. This baseline provides a conservative lower bound for model performance. As shown in Figure~\ref{fig:rmse}, both SeaCast and MedFS significantly outperform persistence across all variables. Notably, SeaCast demonstrates improved forecast skill relative to MedFS, and the gap between the two tends to increase at extended lead times.

\begin{figure*}[h]
\centering
\includegraphics[width=\textwidth]{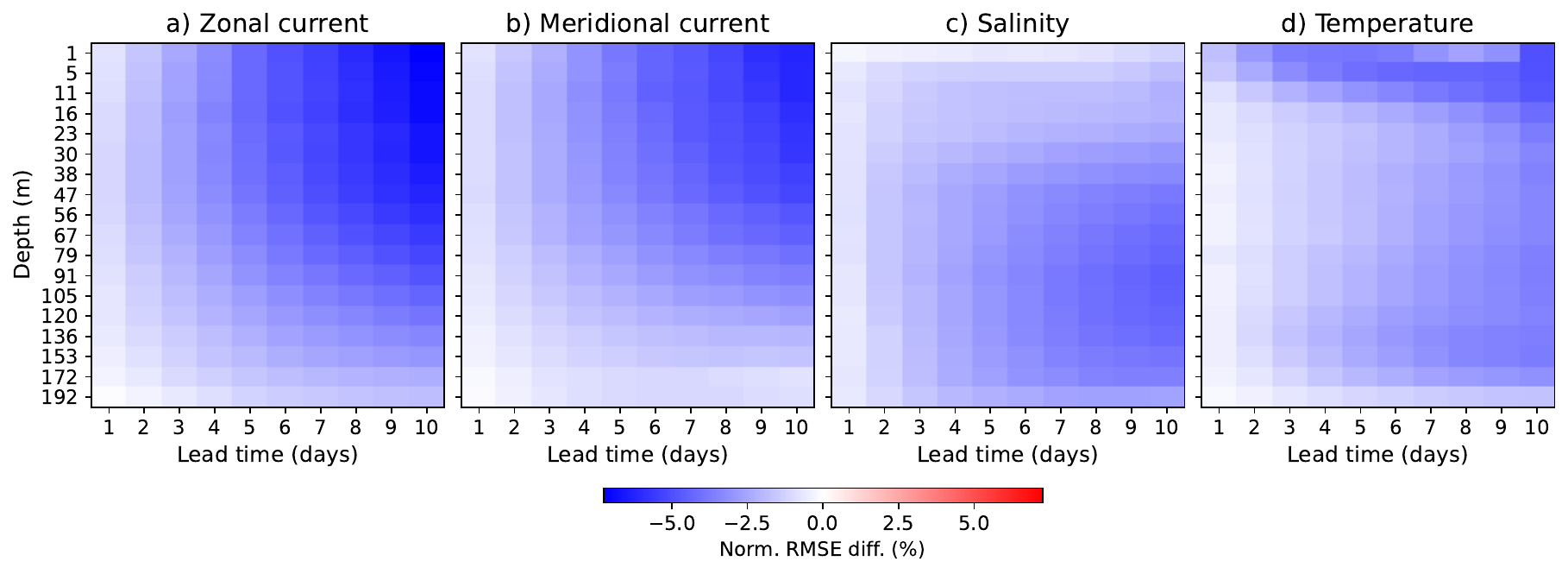}
\caption{Normalized RMSE difference between SeaCast and MedFS across all depth levels, evaluated against analysis fields. Negative values indicate better performance by SeaCast. Improvements are most pronounced near the surface for temperature and the current components, and at greater depths for salinity.}
\label{fig:seacast_vs_medfs}
\end{figure*}

Figure~\ref{fig:seacast_vs_medfs} presents a depth-resolved comparison of SeaCast and MedFS, showing the normalized RMSE difference, defined as $(\mathrm{RMSE}_{\mathrm{SeaCast}}-\mathrm{RMSE}_{\mathrm{MedFS}})/\mathrm{RMSE}_{\mathrm{MedFS}}$, across all vertical levels and lead times. Negative values indicate improved performance by SeaCast. The improvements are generally larger at longer forecast lead times, which is in line with results for the global ocean~\cite{cui2025forecasting}. For temperature and current components, the relative gains are strongest near the surface, while for salinity the largest improvements are found at greater depths. SeaCast generally does not outperform MedFS that much at the lowest level included here (192\,m depth), likely because influence from below this level are unaccounted for and affect the conditions there.

\begin{figure*}[ht]
\centering
\includegraphics[width=\textwidth]{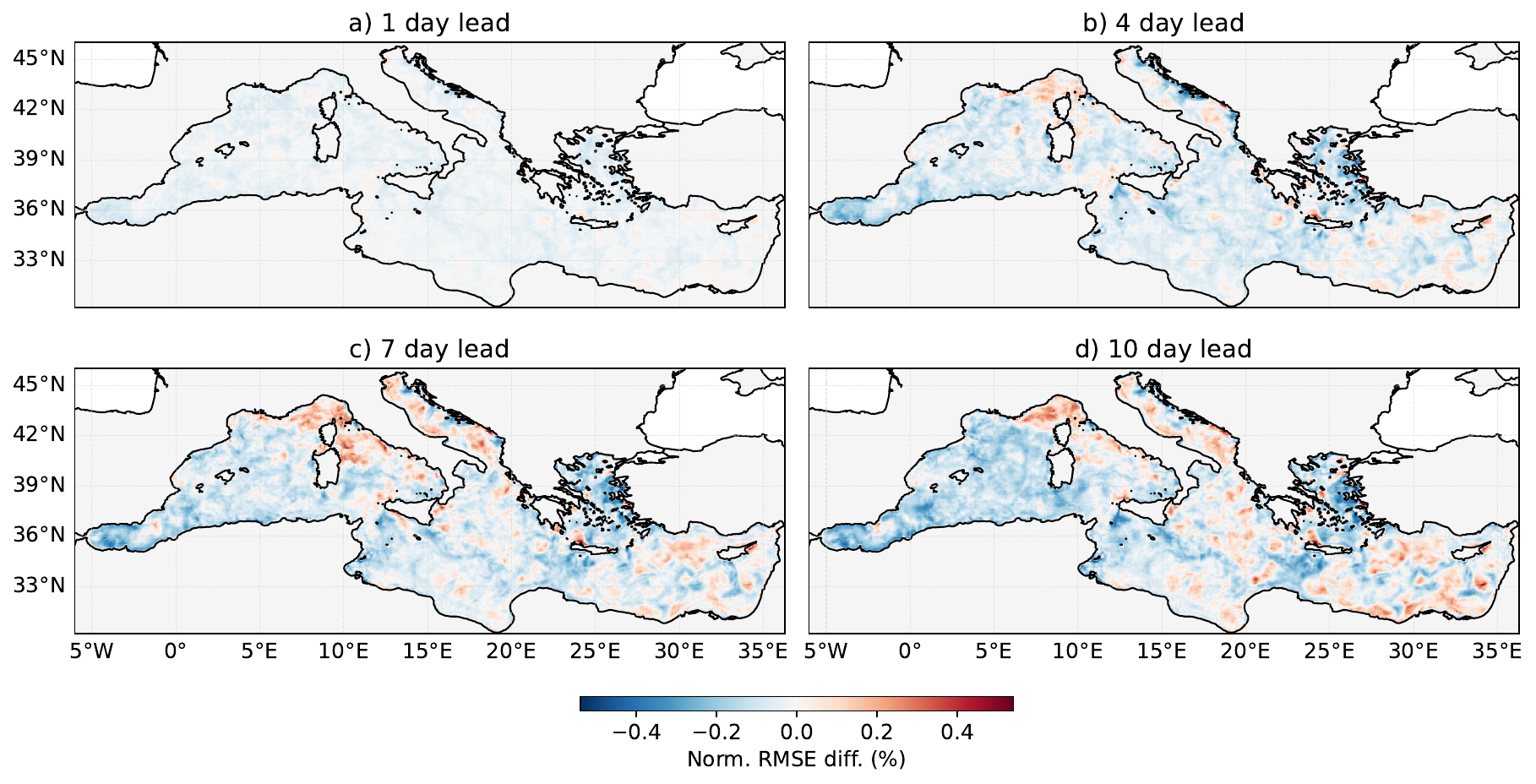}
\caption{Spatial distribution of normalized RMSE difference in SST between SeaCast and MedFS, evaluated against L3 satellite data. Results are shown for lead times of 1, 4, 7, and 10 days. Negative values indicate less error by SeaCast. Relative skill increases with lead time across most of the Mediterranean.}
\label{fig:sst}
\end{figure*}

To further assess regional performance, Figure~\ref{fig:sst} shows the spatial distribution of normalized RMSE differences in SST between SeaCast and MedFS, for forecast lead times of 1, 4, 7, and 10 days. Across the majority of the Mediterranean basin, SeaCast exhibits increasing skill relative to MedFS as the lead time grows. These improvements are particularly notable in the Western basin, especially in the Alboran Sea, and in the Aegean Sea, areas characterized by intense mesoscale processes. Meanwhile, MedFS demonstrates enhanced skill in the Adriatic and Ligurian Seas. This heterogeneous spatial pattern of skill may be linked to differences in the atmospheric forcing used by the two systems.

Additional evaluation metrics and results are presented in Supplementary~C and D, respectively. This includes a dedicated comparison against sparser in-situ observations (Supplementary~D.1), depth-resolved RMSE relative to the analysis fields (Supplementary~D.2), and vertical error profiles (Supplementary~D.3). Qualitative examples of SeaCast forecasts are shown in Supplementary~E.

\subsection{Detecting high temperatures}

Machine learning–based forecasting methods that optimize the mean squared error (MSE) such as we do are trained to predict the expected value of the forecast distribution. As a result, they could potentially struggle with accurately capturing extreme events. In the marine context, one such extreme is unusually warm SST.

To evaluate SeaCast’s ability to predict temperature extremes, we draw inspiration from the marine heatwave (MHW) definition proposed by Hobday et al.~\cite{hobday2016mhw}, where an event is defined as SST exceeding the 90th percentile of the daily climatology for at least five consecutive days, based on an 11-day rolling mean. In our case, we adopt a simplified criterion by identifying individual days when the daily SST exceeds the 90th percentile threshold, allowing us to produce high temperature detections rates at each of the 15 lead times. The SST climatology is computed from L3 satellite observations spanning every day in the years 2008 to 2023, using an 11-day rolling mean. For each calendar day, the 90th percentile is calculated to define the threshold for extreme temperature events.

\begin{figure}[h]
\centering
\includegraphics[width=.4\columnwidth]{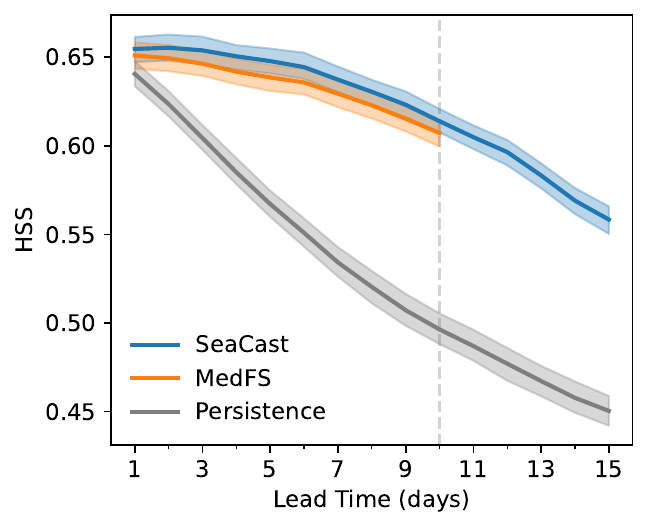}
\caption{Heidke Skill Score (HSS) for detecting SST anomalies above the 90th percentile, relative to climatology. Results are shown for SeaCast, MedFS, and the persistence baseline. The shading corresponds to the 50\% confidence intervals estimated via bootstrapping.}
\label{fig:hss}
\end{figure}

We then evaluate each model’s performance in detecting high temperatures using the \mbox{Heidke} Skill Score (HSS) at each forecast lead time, and estimate 50\% confidence intervals via bootstrapping. Higher HSS values indicate more accurate classification of events relative to random chance. As shown in Figure~\ref{fig:hss}, both SeaCast and MedFS significantly outperform the persistence baseline. Furthermore, SeaCast slightly outperforms MedFS also on temperature extremes. An additional advantage of SeaCast is its ability to provide forecasts up to 15 days, compared to the 10-day horizon of MedFS, offering potential for earlier warnings of MHWs.

\subsection{Effect of atmospheric forcing}
\label{sec:atm_forcing}

Atmospheric forcing plays a critical role in driving ocean dynamics, particularly near the sea surface. To assess the sensitivity of SeaCast to different atmospheric inputs, we perform a controlled ablation experiment by randomly permuting each forcing variable across the spatial grid dimension during inference. This preserves the statistical distribution of the variable but removes its spatial coherence, effectively making it uninformative while leaving all other inputs intact. The resulting performance degradation reveals the relative importance of each atmospheric variable for ocean forecasting.

\begin{figure*}[ht]
\centering
\includegraphics[width=\textwidth]{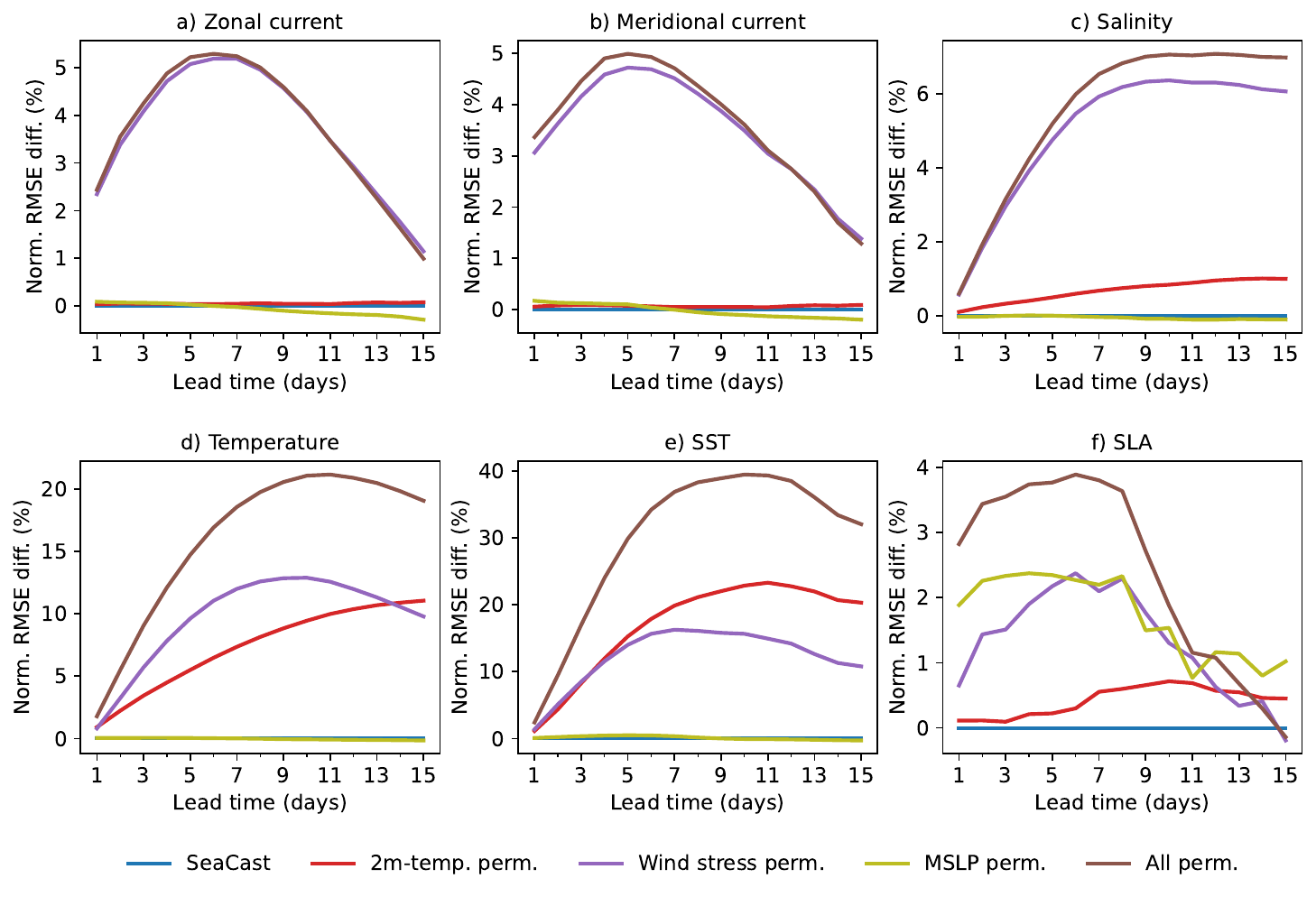}
\caption{Impact of permuted atmospheric forcings on forecast skill. Normalized RMSE difference per lead time is shown relative to the original SeaCast configuration with unmodified atmospheric inputs included as a reference line in blue. Each panel shows a different predicted ocean variable, while the curves correspond to the permuted atmospheric forcing variables. Larger positive values indicate greater importance of the corresponding variable.}
\label{fig:norm_rmse_diff_forcing}
\end{figure*}

Figure~\ref{fig:norm_rmse_diff_forcing} presents the normalized RMSE difference per lead time for each variable, evaluated against the original, unperturbed SeaCast model. The results indicate that wind stress is one of the most critical drivers across all ocean state variables. By transferring momentum from the atmosphere to the ocean, wind stress generates horizontal currents and drives vertical transport. Thus, the results show that for zonal and meridional currents in particular, wind stress is the sole contributor to predictive skill among the atmospheric forcing components.

Salinity forecasts are also strongly influenced by wind stress, an intuitive result given its central role in driving surface mixing and the lateral advection of freshwater. The second most influential variable for salinity is the 2-meter air temperature, which indirectly affects surface buoyancy and stratification triggering convective overturning and mixing fresh or salty surface water into the subsurface~\cite{marshall1999open}. For temperature profiles, wind stress again dominates across depth. Near the surface, however, 2-meter temperature becomes increasingly important due to its direct influence on air–sea heat flux, particularly through modulation of the sensible heat flux and its impact on vertical mixing in the subsurface~\cite{large1994oceanic}.

For SLA, the dominant atmospheric driver is MSLP, which directly affects the sea level through the inverted barometer effect~\cite{wunsch1997atmospheric} and large-scale ocean response. Moreover, MSLP gradients give rise to surface wind stress through atmospheric geostrophic balance~\cite{gill1982atmosphere}, making wind stress the second most significant contributor to SLA.

The patterns we see here confirm known ocean–atmosphere coupling mechanisms and support the robustness of SeaCast's sensitivity to physically meaningful drivers. A depth-resolved analysis of the effect of atmospheric forcing on individual variables is further provided in Supplementary~D.2.

\begin{figure*}[ht]
\centering
\includegraphics[width=\textwidth]{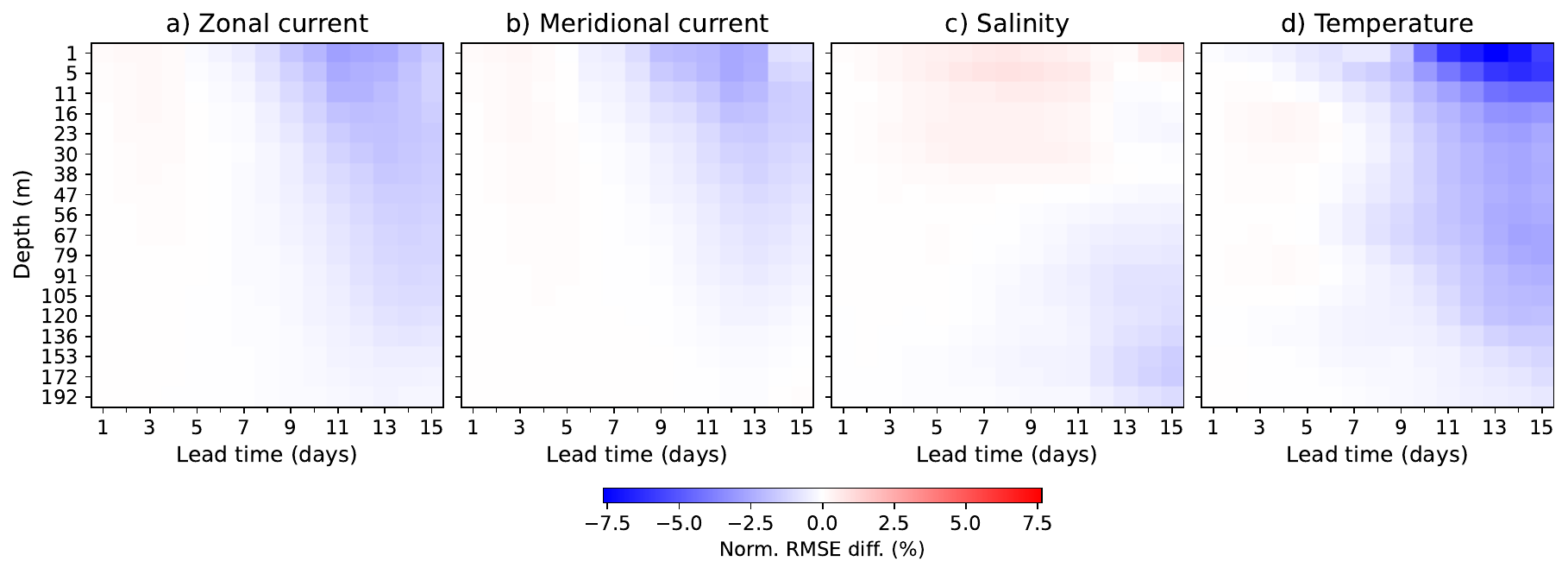}
\caption{Normalized RMSE difference scorecard comparing SeaCast performance when forced with AIFS versus ENS atmospheric inputs. Evaluations are performed against analysis fields across all depth levels. Negative values indicate less error when using AIFS forcing.}
\label{fig:aifs_vs_ens}
\end{figure*}

Additionally, we assess SeaCast’s performance using two different atmospheric forcing products: AIFS and ENS. Both forcings include the same set of variables, providing a consistent basis for comparison. Figure~\ref{fig:aifs_vs_ens} presents scorecards of normalized RMSE differences across depth levels. Results indicate that the AIFS forcing leads to improved forecast skill for currents and temperature, particularly at longer lead times, with the most notable gains observed in SST. This is consistent with the evaluation of AIFS itself, which demonstrates lower forecast errors compared to its numerical counterpart beyond day one~\cite{lang2024aifs}. However, for salinity, SeaCast with ENS forcing exhibits slightly better performance in the upper layers here.

\subsection{Effect of training period}

\begin{figure*}[ht]
\centering
\includegraphics[width=\textwidth]{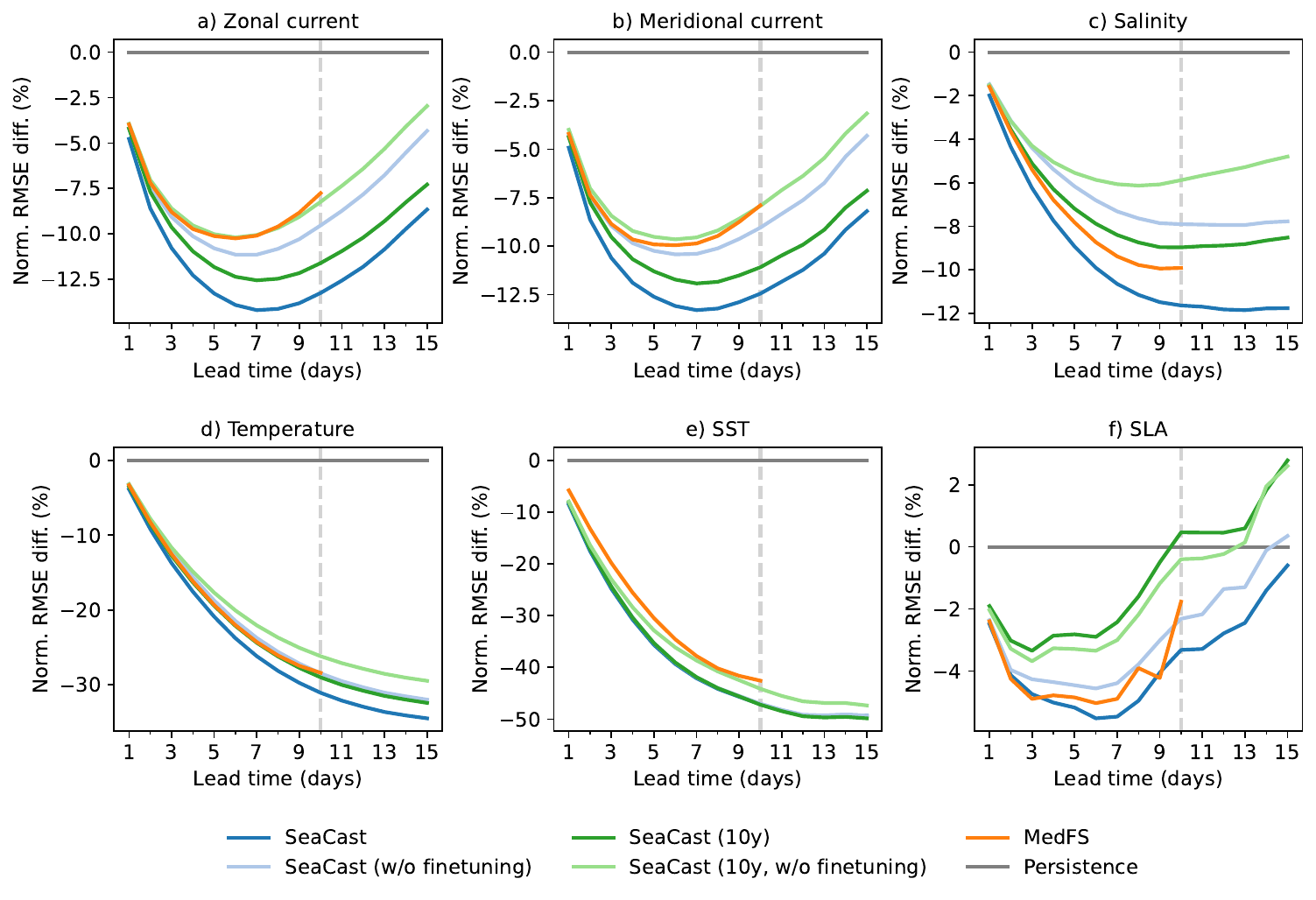}
\caption{Normalized RMSE difference per lead time for SeaCast variants and MedFS compared to a persistence baseline. SeaCast is trained on 35 years of reanalysis (1987--2021) with 2-year fine-tuning (2022--2023). The 10y variant uses only 8 years of reanalysis (2014--2021), with or without the same fine-tuning. Results show the impact of training duration and fine-tuning on forecast performance.}
\label{fig:norm_rmse_diff}
\end{figure*}

To evaluate the impact of training data duration and fine-tuning on model performance, we compare SeaCast variants trained on different historical timespans and examine the effect of omitting fine-tuning. Figure~\ref{fig:norm_rmse_diff} presents the normalized RMSE difference relative to a persistence baseline across forecast lead times, with MedFS included for reference. The full SeaCast model is trained on 35 years of reanalysis data (January 1987 to December 2021), followed by fine-tuning on 2 years of more recent analysis data (January 2022 to December 2023). We compare this to a version trained solely on the 35-year reanalysis dataset, without fine-tuning. Additionally, we include a shorter-term model variant trained on 8 years of reanalysis data (January 2014 to December 2021), both with and without fine-tuning on the same 2-year analysis period. This shorter-term model is denoted as SeaCast (10y). The models are trained using the same number of epochs and follow identical epoch-wise learning rate schedules. As a result, SeaCast (10y, w/o finetuning) undergoes fewer optimization steps during the pre-training phase compared to the full model, and requires a smaller computational budget (6\,h vs. 20.5\,h on 64 GPUs).

The results show that the 10-year variants of SeaCast generally perform on par with or slightly better than MedFS for zonal and meridional currents, temperature profiles, and sea surface temperature. However, for salinity and SLA, only the full SeaCast model, trained on 35 years of reanalysis data and fine-tuned on analysis data, outperforms MedFS. The SLA error for the fine-tuned 10-year variant is slightly higher than that of the base version, which may seem counterintuitive. This can be partly explained by the fact that SLA evaluation is performed along satellite tracks, whereas training is done on regularly gridded simulation outputs. Minimizing MSE on simulated fields does not guarantee lower error on satellite tracks, even if SLA is assimilated into the physics-based model. Moreover, the satellite tracks yield a sparser and more variable target compared to other fields, producing a wider confidence interval, as observed in Figure~\ref{fig:rmse}. On other fields, the fine-tuned model consistently outperforms the base version. These findings are encouraging for regional settings with access to limited historical data, suggesting that even 10 years of reanalysis data and a more modest compute budget can produce forecast skill competitive with numerical forecast models.

Fine-tuning on analysis data yields consistent performance improvements across variables, which is expected given that several evaluation targets are derived from the same analysis system. These analysis states are more up-to-date than the reanalysis and incorporate system enhancements not present in the older data, including open boundary conditions at the Dardanelles Strait, coupling with a wave model~\cite{clementi2017coupling}, and tidal forcing~\cite{mcdonagh2024characteristics}. By fine-tuning on this updated dataset, SeaCast adapts to the characteristics of the current operational system, leading to better alignment with the evaluation targets and improved overall forecast accuracy.

\section{Discussion}
\label{sec:discussion}

In this study, we introduced SeaCast, a graph-based machine learning model designed for high-resolution regional ocean forecasting. SeaCast learns directly from historical reanalysis and analysis data to predict key ocean variables in the Mediterranean Sea. The model demonstrates improved skill for all modeled variables across all depth levels compared to the operational MedFS when considering both analysis fields and satellite observations as references. Another benefit of SeaCast is its speed: once trained, it delivers a full 15‑day, 18‑level forecast on a $1/24^\circ$ grid in just 20\,s on a single GPU, substantially faster than the physics-based model on a CPU cluster. This performance can enable new capabilities for Mediterranean forecasting such as rapid scenario testing and ensemble forecasting with many members.

Our sensitivity experiments underscore the central role of atmospheric forcing, where wind stress emerges as a dominant driver of predictive skill across all ocean variables, with surface temperature and MSLP exerting secondary influence motivated by the the air–sea heat‑flux and inverted‑barometer effects. These results suggest that physically meaningful coupling mechanisms are captured by the data‑driven framework. We further demonstrate that the length and recency of the training period materially affect forecast accuracy. The 10‑year variant of SeaCast achieves performance on par with MedFS for currents, temperature, and SST, while relying on very attainable historical data archives and computational resources.

Despite the promising outcomes, several avenues remain for future improvement. First, enhancements to the underlying reanalysis systems, in line with recent progress on MedFS~\cite{coppini2023forecast}, could further increase the fidelity of data-driven forecasting models. Additionally, increasing the temporal resolution of archived datasets, e.g. from daily to 6-hourly fields, would provide richer training data and enable models to capture diurnal variability. Furthermore, recent work has demonstrated that machine learning can be used to produce continuous-time forecasts~\cite{ruhling2023dyffusion, andrae2025continuous}, a concept that could potentially be adapted to ocean modeling.

While SeaCast takes as input lateral boundary forcing from MedFS, leveraging its existing regridded connection to the global ocean, early studies in limited area weather modeling with machine learning provide interesting alternative ways for handling boundary conditions, such as using a globally stretched grid~\cite{nipen2024regional} or learned mappings from surrounding forecasts~\cite{adamov2025building}. Integrating both data-driven global ocean and atmospheric forecasts, which can be rolled out arbitrarily far, into machine learning-based regional ocean models would enable similarly long lead-time predictions for these models as well. This opens up the question of how far into the future such forecasts can remain skillful.

The interdependence of physics, biogeochemistry, and wave dynamics in the MedFS system~\cite{coppini2023forecast} suggests a natural extension for machine learning frameworks: a coupled neural architecture that jointly predicts sea state, biochemical tracers, and wave variables, enabling truly holistic forecasts. While these particular components are usually represented on compatible grids, recent advances in foundation models for weather and climate have shown that even heterogeneous spatial and temporal datasets, can be effectively used for training, resulting in models that generalize across diverse prediction tasks~\cite{nguyen2023climax, bodnar2025foundation}. Adopting this paradigm for ocean forecasting could enable the development of transferable representations, supporting scalable and adaptable forecasting systems across regions and variable types.

Another direction of future work is developing probabilistic ocean forecasting systems. While SeaCast currently produces deterministic forecasts akin to traditional operational ocean forecasting systems, emerging generative machine learning methods for weather forecasting emphasize uncertainty quantification through ensemble generation~\cite{price2025probabilistic, larsson2025diffusion}. These approaches enable better decision-making by accounting for forecast spread and producing potentially more physically consistent ensemble members, which could be very useful for risk-sensitive applications such as early warning systems and coastal hazard planning.

In conclusion, SeaCast demonstrates the viability of machine learning for regional ocean forecasting by outperforming a physics‑based system. This work also presents novel sensitivity experiments in data‑driven ocean forecasting, revealing the roles of atmospheric forcing components and training period characteristics. Further improvements in data fidelity, model coupling, and probabilistic prediction will be key to realizing the full potential of machine learning–based regional ocean forecasting.

\section{Methods}
\label{sec:methods}

\paragraph{Problem definition}
The forecasting problem is characterized by mapping a sequence of initial states $X^{-p:0} = (X^{-p}, ..., X^{0})$, where $p$ is the length of the past window, to a sequence of future states $X^{1:T} = (X^{1}, \dots, X^{T})$, where $T$ is the length of the forecast. Each state $X^t \in \mathbb{R}^{N \times d_x}$, contains $d_x$ variables at $N$ locations represented on a grid. Variables include both those at multiple vertical levels and single-level surface measures. In addition to this, forcing inputs $F^{1:T} = (F^{1}, ..., F^{T})$ encompassing known dynamic factors relevant to the forecasting problem, are also available.

\subsection{SeaCast}
\label{sec:seacast}

\paragraph{Graph-based neural forecasting}
Initial states typically cover the two preceding time steps $X^{-1:0}$, enabling the capture of first-order dynamics~\cite{lam2023graphcast}. Concatenating a single-step forecast $\hat{X}^{t} = f(X^{t-2:t-1}, F^{t})$ to the initial state and repeating the process with $t$ incremented by one allows autoregressive forecasts $\hat{X}^{1:T}$ of arbitrary length $T$. An integral part of this framework is mapping from $N$ surface grid points onto a mesh graph $\mathcal{G}_M = (\mathcal{V}_M, \mathcal{E}_M)$ coarser than the simulation domain~\cite{fortunato2022multiscale}. The function $f$ is implemented as a sequence of GNN layers following an encode-process-decode architecture~\cite{sanchez2020learning} where: 1) grid inputs are encoded onto the mesh representation; 2) a number of GNN layers process this latent representation; 3) the processed data is mapped back onto the original grid. SeaCast predicts the next step as a residual update to the most recent input state, making it an easier learning task than predicting the next state directly. The mappings between grid and mesh nodes occur through bipartite grid-to-mesh $\mathcal{G}_{G2M}$ and mesh-to-grid $\mathcal{G}_{M2G}$ graphs. Upwards node and edge updates are facilitated through GNN layers in the form of \emph{propagation networks}~\cite{oskarsson2024probabilistic}, and the remaining updates through \emph{interaction networks}~\cite{battaglia2016interactionnet}, with all layers mapping to a latent dimensionality $d_z$. This design leverages the inherent inductive bias of interaction networks to retain information, whereas propagation networks are more effective at forwarding new information through the graph. Multilayer perceptrons within GNN layers consists of a single hidden layer using the Swish activation function~\cite{ramachandran2017swish}, followed by layer normalization~\cite{ba2016layernorm}.

\paragraph{Hierarchical graph}

The hierarchical mesh structure described in \cite{oskarsson2024probabilistic} consists of multiple graph levels $\mathcal{G}_1, \dots, \mathcal{G}_H$, where each level $\mathcal{G}_h = (\mathcal{V}_h, \mathcal{E}_h)$  gets progressively coarser. The first level connects directly to the grid, forming $\mathcal{G}_{G2M} = (\mathcal{V}_G \cup \mathcal{V}_1, \mathcal{E}_{G2M})$ and $\mathcal{G}_{M2G} = (\mathcal{V}_G \cup \mathcal{V}_1, \mathcal{E}_{M2G})$. A processing step on the mesh is defined as a complete sweep through the hierarchy $\mathcal{G}_{h, h+1} = (\mathcal{V}_h \cup \mathcal{V}_{h+1}, \mathcal{E}_{h, h+1})$ where the graph sequences are $\mathcal{G}_{1,2}, ..., \mathcal{G}_{H-1,H}$ and $\mathcal{G}_{H,H-1}, ..., \mathcal{G}_{2,1}$, respectively.

\paragraph{Mesh construction} Regional oceans can take very irregular shapes, calling for a customized mesh for the modeled geographical area. The foundation for our mesh is a quadrilateral construction used for graph-based limited area weather modeling~\cite{oskarsson2024probabilistic}. It is initialized by selecting only the nodes corresponding to the ocean surface grid. All nodes are connected with bidirectional edges to their neighbors horizontally, vertically and diagonally. Nodes on higher resolution levels are positioned in the center of a $3 \times 3$ square on the level below. Upward edges are created by connecting each node at level $h$ to the closest nodes at level $h+1$, and the downward edges mirror these. Edges crossing land areas with a threshold of 8 grid points are excluded, both for inter- and intra-level graphs. This procedure results in a mesh that conforms to the shape of the regional ocean.

\paragraph{Rollout masking}

We want to ensure 1) that the learning task of the model is exclusively for grid nodes inside the regional ocean at each depth level, and 2) that the predictions are aligned with the influence of the global ocean. To address the first point we only propagate predictions part of the internal depth-wise ocean grid $\mathbb{G}$ in the autoregressive rollout. In response to the second point, boundary forcing is applied at each time step by replacing predictions inside the boundary region $\mathbb{B}$ with the ground truth forecast $X_t$. We update the row for each node $v$ as:
\begin{equation}
    \hat{X}^{t}_{v,i} \leftarrow \left(\mathbb{I}_{\{v \in \mathbb{G}_{l,i}\}} - \mathbb{I}_{\{v \in \mathbb{B}_{l,i}\}}\right) \hat{X}^{t}_{v,i} + \mathbb{I}_{\{v \in \mathbb{B}_{l,i}\}} X^{t}_{v,i}\ \forall l \in \mathbb{L}_i
\end{equation}  
Here, $\mathbb{I}_{\{\cdot\}}$ denotes the indicator function, equal to 1 if the specified condition is true and 0 otherwise. $\mathbb{G}_{l,i}$ represents the set of oceanic grid nodes at depth level $l$ associated with variable $i$, with $\mathbb{B}_{l,i}$ defined analogously for the boundary region. Finally, $\mathbb{L}_i$ denotes the set of vertical levels associated with variable $i$.

\paragraph{Training objective}

The model is trained to minimize the mean squared error (MSE) between the predictions and the ground truth over a rolled-out sequence of states. The loss function we use is similar to what is commonly applied in MLWP, with the distinction that we account for the bathymetry, or the ocean grid structure at each depth level. The complete loss function is defined as:
\begin{equation}
    \mathcal{L} = \frac{1}{T_{\text{rollout}}} \sum_{t=1}^{T_{\text{rollout}}} \sum_{i=1}^{C} \lambda_i \sum_{l=1}^{L_i} \frac{w_l}{N_l} \sum_{v=1}^{N_l} a_v \left( \hat{X}^{t}_{v, i} - X^{t}_{v, i} \right)^2
\end{equation}
where $T_{\text{rollout}}$ is the number of steps in the rollout, $C$ is the number of ocean variables in the tensor, $L_i$ is the number of depth levels for variable $i$, $N_l$ is the the number of ocean grid nodes at depth level $l$, $a_v$ is the latitude-longitude area of grid cell $v$ normalized to unit mean, $w_l$ is the loss weight for depth level $l$, and $\lambda_i$ is the inverse variance of time differences for variable $i$. 

The area factor $a_v$ accounts for variations in the size of grid cells due to the spherical geometry of the Earth, assigning lower weights to cells near the poles and higher weights near the equator, and is normalized to have unit mean across the domain. The depth-specific weight $w_l$ is designed to prioritize prediction accuracy in the upper ocean, reflecting the greater societal relevance of shallower depths. For depth-resolved variables, $w_l$ is proportional to $(200 - d)$, where $d$ is the depth in meters, and the weights are normalized to have unit mean across all depths. The single-level SSH variable is assigned a weight of one. Further normalizing by the magnitude of the dynamics for each feature, signified by the $\lambda_i$ factor, ensures that the model evaluates errors across vertical levels on a physically meaningful scale~\cite{keisler2022forecasting}.

\paragraph{Model and training}

We train SeaCast with 3 mesh levels and 3 processor layers with $d_z = 256$ hidden units, totaling $17.7$\,M trainable parameters. Training initiates with a warm-up phase of five epochs, starting from a learning rate of $10^{-5}$ and incrementing epoch-wise to a base rate of $10^{-3}$. Following the warm-up, we employ a cosine decay schedule. For optimization, we use AdamW~\cite{loshchilov2019adamw}, configured with $\beta_1 = 0.9$, $\beta_2 = 0.95$, and a weight decay of $0.1$. SeaCast is trained on reanalysis data for 200 epochs using a local batch size of 1. The model is further finetuned on analysis data for 30 epochs, with the same learning rate schedule, but the initial learning rate set to $10^{-7}$ and the base rate as $10^{-5}$. Furthermore, the number of rollout steps is progressively increased to 2 at epoch 10 and 3 at epoch 20. Additional rollout steps can be added to the loss function, although this comes at an increased training cost, and have been observed to yield diminishing returns in MLWP~\cite{keisler2022forecasting}. The pretraining on 35 years of daily reanalysis data took 20.5\,h on 64 AMD MI250x GPUs (1312 GPU hours) in a data-parallel configuration, and finetuning on 2 years of analysis data took 3.5\,h on 8 GPUs (28 GPU hours).

\paragraph{Computational complexity}

SeaCast can produce a complete 15-day forecast in 20 seconds on a single AMD MI250x GPU, which is roughly equivalent to 1.3 seconds per simulated day ($\sim 178$~SYPD). The SeaCast forecast includes 18 depth levels, and outputs predictions at a daily temporal resolution. In contrast, the computational time required for the MedFS is approximately 70 minutes to run a 10-day forecast ($\sim 0.6$~SYPD), using 89 CPU cores. This includes the time for generating and writing outputs for all 141 vertical levels. Both models produce outputs at the same 1/24° spatial resolution.

\subsection{Dataset}

\paragraph{Ocean state} The Mediterranean Sea physics analysis and forecasting system leverages a two-way coupled hydrodynamic–wave modeling framework composed of the Nucleus for European Modelling of the Ocean (NEMO) v4.2~\cite{madec2017nemo}, which includes an explicit representation of tides, and WAVEWATCH III v6.07~\cite{ww3dg2019wavewatch}. Model solutions are corrected using the 3D variational data assimilation scheme OceanVar~\cite{dobricic2008oceanvar}, which integrates quality-controlled in situ and satellite observations to improve ocean dynamics. The Mediterranean reanalysis system is based on NEMO v3.6 (without tides) and OceanVar, assimilating reprocessed observational data. The model is implemented at a high horizontal grid resolution of 1/24°, utilizing a total of 141 vertical levels with uneven spacing. By choosing every other depth until 200\,m the list of selected depths included in this study becomes: 1.02, 5.46, 10.5, 16.3, 22.7, 29.9, 37.9, 46.7, 56.3, 66.9, 78.6, 91.2, 105, 120, 136, 153, 172, and 192 meters. The topography is based on an interpolation of the General Bathymetric Chart of the Oceans (GEBCO) 30 arcsecond grid~\cite{weatherall2015bathymetry}. The test data consists of daily forecasts issued from July 3rd, 2024 to December 31st, 2024. The last date used for evaluation is hence January 14th, 2025, as SeaCast provides 15-day forecasts.

\paragraph{Atmospheric forcing}

Atmospheric forcing play an important role in data-driven modeling of the ocean's response to atmospheric conditions~\cite{subel2024building}, especially in driving marine dynamics near the sea surface. We incorporate four key atmospheric variables: the 2-meter temperature, MSLP, and the 10-meter zonal and meridional wind stress components derived from wind components using the Hellerman and Rosenstein formulation~\cite{hellerman1983windstress} for the drag coefficient. The atmospheric data are sourced from daily mean aggregates of 6-hourly ERA5 reanalysis data~\cite{hersbach2023era5}. For testing, we compare the 6-hourly daily means of the numerical ENS and the data-driven AIFS forecasts. All atmospheric forcing variables are bi-linearly interpolated from their native 1/4° resolution to the 1/24° resolution of the sea grid. Additionally, the sine and cosine of the day of year, normalized between 0 and 1, are included as forcing features to account for seasonal variations. The atmospheric forcing used in our model is windowed over three consecutive time steps. This means that each forcing input $F_t$ includes data from times $t - 1$, $t$, and $t + 1$.
 
\paragraph{Lateral boundary forcing}

We define the lateral boundary as the grid nodes west of longitude -5.2° to the edge the grid, covering the Strait of Gibraltar, as well as the Dardanelles Strait, which spans latitudes $39.9^\circ$ to $40.4^\circ$ and longitudes $25.9^\circ$ to $26.4^\circ$. Note that we use a boundary region that lies inside the propagated data grid, allowing us to use Mediterrenaen forecast data as boundary forcing. Current ocean forecasts at CMEMS are available 10 days in the future for the most part, following the length of HRES atmospheric forcing. However, we use the ENS/AIFS standard of 15-day forecasts. Hence we have to increase the length of the boundary forcing, and we do so by persisting the last forecast state in the boundary region five times at the end.

\paragraph{Satellite data}
\label{sec:satellite_data}

Satellite observations of SST and SLA over the Mediterranean Sea are used for evaluation. For SST, we use the super-collated Level-3S (L3S) multi-sensor merged product~\cite{nardelli2013sstobs} delivered through CMEMS, which provides daily SST fields at a spatial resolution of $1/16^\circ$. These observations are representative of nighttime conditions, thereby minimizing the effects of the diurnal heating cycle. The L3S dataset is constructed by merging only the highest quality Level-2P (L2P) input data from several satellite sensors, selected within a strict local nighttime window to reduce cloud contamination and sensor inconsistencies. The main contributing sensors include SLSTR from Sentinel-3A and -3B, VIIRS from NOAA-20 and Suomi-NPP, AVHRR from Metop-B and -C, and SEVIRI from Meteosat. Model forecasts are bilinearly interpolated to the L3S SST grid during evaluation.

For SLA, we use the CMEMS along-track L3 near-real-time product. This product contains high-resolution measurements, where we use 5\,Hz measurements from satellite altimeters HaiYang-2B, Jason-3 (interleaved orbit), Sentinel-3A and -3B, Sentinel-6A, and the SWOT mission (nadir track). The filtered version of this dataset is used to minimize the influence of measurement noise. Model SSH is converted to SLA and bilinearly interpolated onto the satellite track coordinates during evaluation. The process of aligning model and observation data is detailed in Supplementary~A.3.

\subsection{Evaluation metrics}

We assess the performance of SeaCast using several variations of RMSE, which quantifies the average magnitude of prediction errors relative to the reference observations. RMSE is computed per lead time and per variable, and includes area weighting to account for the changing size of ocean grid cells with latitude, consistent with the loss function used during training. For the 3D variables temperature, salinity, and currents, we report depth-averaged RMSE to provide a vertically integrated summary of forecast skill at each lead time. For SST, we additionally compute per-location RMSE by comparing model outputs interpolated onto the L3S SST grid to the observations.

To evaluate the model’s ability to detect temperature extremes, we use HSS, which measures categorical forecast skill relative to random chance. HSS is computed based on a binary classification of observed threshold exceedances, defined as SST values above the 90th percentile of the daily climatology, using an 11-day rolling mean. An HSS of 1 indicates a perfect forecast, 0 indicates no skill beyond chance, and negative values indicate performance worse than a random forecast.

Additional details on the verification metrics and their mathematical definitions are provided in Supplementary~C.

\section*{Data availability}

The model weights and preprocessed data to reproduce the results for this study are stored at \url{https://doi.org/10.57967/hf/5342}. The original data is publicly available via CMEMS for the ocean data and ECMWF for the atmospheric data. The Mediterranean reanalysis and analysis data can be downloaded from \url{https://doi.org/10.25423/cmcc/medsea_multiyear_phy_006_004_e3r1} and \url{https://doi.org/10.25423/cmcc/medsea_analysisforecast_phy_006_013_eas8}. The atmospheric forcing is obtained from \url{https://doi.org/10.21957/open-data}. The observed SST, SLA, and in-situ data used for evaluation are available via \url{https://doi.org/10.48670/moi-00171}, \url{https://doi.org/10.48670/moi-00140}, and \url{https://doi.org/10.48670/moi-00044}, respectively.

\section*{Code availability}

The source code for SeaCast is available at \url{https://doi.org/10.5281/zenodo.15367632}. The repository contains scripts for data downloading and preprocessing, as well as code for training and evaluating the models presented.

\backmatter


% \bibliography{sn-bibliography}% common bib file

\vspace*{\baselineskip}

\bmhead{Supplementary information}

We provide a Supplementary Materials document that contains additional methodological details, as well as supplementary text and results that supports the main claims of our paper.

\bmhead{Acknowledgements}

This work was financially supported by the Research Council of Finland under the FAISER project (grant no. 361902). Computing resources were provided by the LUMI supercomputer, owned by the EuroHPC Joint Undertaking and hosted by CSC–IT Center for Science.

\bmhead{Declarations}

\begin{itemize}
\item Funding: This project was supported by the Research Council of Finland.
\item Conflict of interest: None to report.
\item Ethics approval and consent to participate: Not applicable.
\item Consent for publication: All authors have reviewed the manuscript and consent to its publication.
\item Data availability: See section on data availability.
\item Materials availability: Not applicable.
\item Code availability: See section on code availability.
\item Author contribution: D.H. conceptualized the project and conducted the experiments.
E.C. and I.E. assisted with the assessment of regional ocean forecasting systems and provided guidance on interpreting the results. T.R. supervised the project. All authors provided feedback on results at various stages of the project and contributed to the writing of the manuscript.
\end{itemize}

\end{document}

\title{Supplementary Materials}
\subtitle{%
  \centerline{Accurate Mediterranean Sea forecasting via graph-based}%
  \centerline{deep learning}%
}
\author{Daniel Holmberg, Emanuela Clementi, Italo Epicoco, Teemu Roos}

\maketitle

\tableofcontents

\clearpage

\section{Data details}
\label{app:data_details}

\subsection{Region considered}

\begin{figure}[h]
\centering
\includegraphics[width=\textwidth]{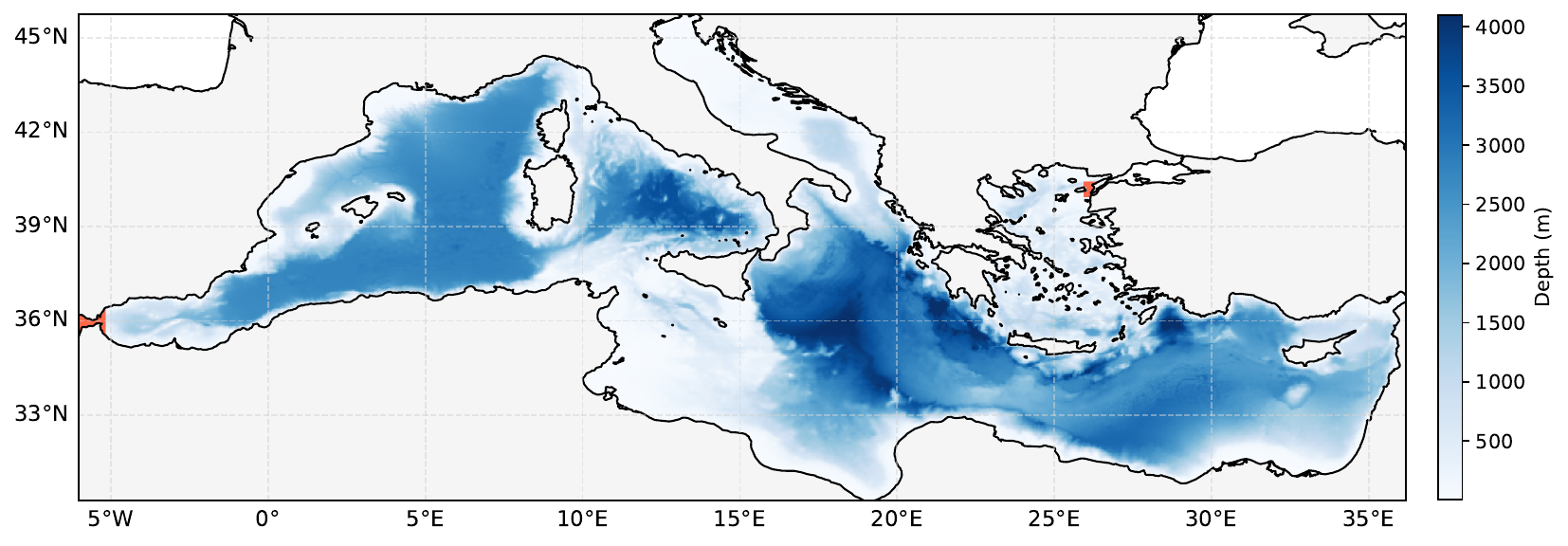}
\caption{Bathymetry of the Mediterranean Sea domain used in SeaCast, with forcing regions highlighted in red. The western boundary at the Strait of Gibraltar and the eastern boundary at the Dardanelles Strait are designated as open boundaries, where boundary conditions are applied to account for water exchange with adjacent seas.}
\label{fig:region}
\end{figure}

The domain used in this study covers the Mediterranean Sea, with horizontal resolution of $1/24^\circ$, from the surface down to 200~m depth, and includes two lateral open boundary regions where external forcing is applied. These boundaries allow the model to account for the inflow and outflow of water with neighboring seas. The first open boundary is located at the western edge of the domain, near the Strait of Gibraltar connecting to the Atlantic Ocean, and includes all grid cells west of longitude $5.2^\circ\mathrm{W}$. The second open boundary is situated in the northeastern corner of the domain, corresponding to the Dardanelles Strait connecting to the Black Sea, and spans the region bounded by latitudes $39.9^\circ\mathrm{N}$ to $40.4^\circ\mathrm{N}$ and longitudes $25.9^\circ\mathrm{E}$ to $26.4^\circ\mathrm{E}$. These regions are highlighted in red in Figure~\ref{fig:region}, overlaid on the bathymetry map of the Mediterranean Sea.

The minimum rectangle in which the Mediterranean Sea fits at the current resolution is 371 by 1013 grid cells. This is a total of 375 823 cells, whereas the actual sea surface of the Mediterranean only has $N$ = 144 990 cells, or 45\% of the total number. Only this subset of grid cells has to be processed by SeaCast. The complete data grid is shown in Figure~\ref{fig:region}. Single-level features, forcing features, and static fields all cover the surface area, whereas the simulated currents, temperature and salinity have values on several vertical levels.

\subsection{Variables considered}

The sea physics dataset used to train and evaluate SeaCast consists of dynamic variables, static fields, and atmospheric forcing inputs. Table~\ref{tab:variables} provides a detailed overview of all included features. The core variables represent the three-dimensional state of the ocean, including currents, temperature, salinity provided at 18 vertical levels, as well as SSH (73 predicted fields in total). The bathymetry and mean dynamic topography static fields are used to define the spatial structure of the ocean domain. Finally, surface atmospheric forcing variables including wind stress, 2-meter temperature, and MSLP are incorporated to drive the ocean dynamics, along with cyclical encodings of time of year to account for seasonal variability.

\begin{table}[h]
\centering
\caption{Summary of all variables, static fields, and forcing features in the Mediterranean Sea physics dataset.}
\begin{tabular}{lll}
\toprule & Unit & Vertical Level \\
\midrule
Variables & & \\
\midrule
Zonal sea water velocity & m/s & 18 depths \\
Meridional sea water velocity & m/s & 18 depths \\
Sea water salinity & psu & 18 depths \\
Sea water potential temperature & \textdegree C & 18 depths \\
Sea surface height above geoid & m & Sea surface \\
\midrule
Static fields & & \\
\midrule
Sea floor depth below geoid & m & Sea floor \\
Mean dynamic topography & m & Sea surface \\
Latitude & \textdegree & - \\
Longitude & \textdegree & - \\
\midrule
Forcing & & \\
\midrule
Zonal wind stress & Pa & Sea surface \\
Meridional wind stress & Pa & Sea surface \\
2-meter temperature & \textdegree C & Sea surface \\
Mean sea level pressure & Pa & Sea surface \\
Sine of time of year & - & - \\
Cosine of time of year & - & - \\
\bottomrule
\end{tabular}
\label{tab:variables}
\end{table}

\subsection{Satellite data}

Predicted SST and SLA are compared also to satellite observations. SST is fairly straightforward to compare on the satellite L3S SST grid, but for SLA there is a bit of a procedure to align it with L3 measurements. To compute the observed SLA used in evaluation, we take the filtered SLA satellite data and apply dynamic atmospheric correction, ocean tide correction, and internal tide correction. These additional terms account for physical processes that influence sea level but are not captured directly in the raw altimeter measurements.

In contrast, the sea level forecasts from MedFS and SeaCast are provided as absolute SSH relative to the geoid. To make this comparable to observed SLA, we subtract the model’s mean dynamic topography, effectively aligning the model output to a time-mean reference sea level similar to the satellite product, which uses the 1993–2012 period as its climatological baseline.

Each satellite track is split into ocean-only segments by applying a sea mask. Tracks are divided wherever land is encountered, and only segments containing a minimum number of four valid ocean points are retained for evaluation. Finally, to minimize any residual bias in comparisons along individual satellite tracks, we remove the average SLA value from each track in both the model forecasts and the observations.

\clearpage
\section{Model Details}
\label{app:model_details}

\begin{figure}[h]
    \centering
    \includegraphics[width=\textwidth]{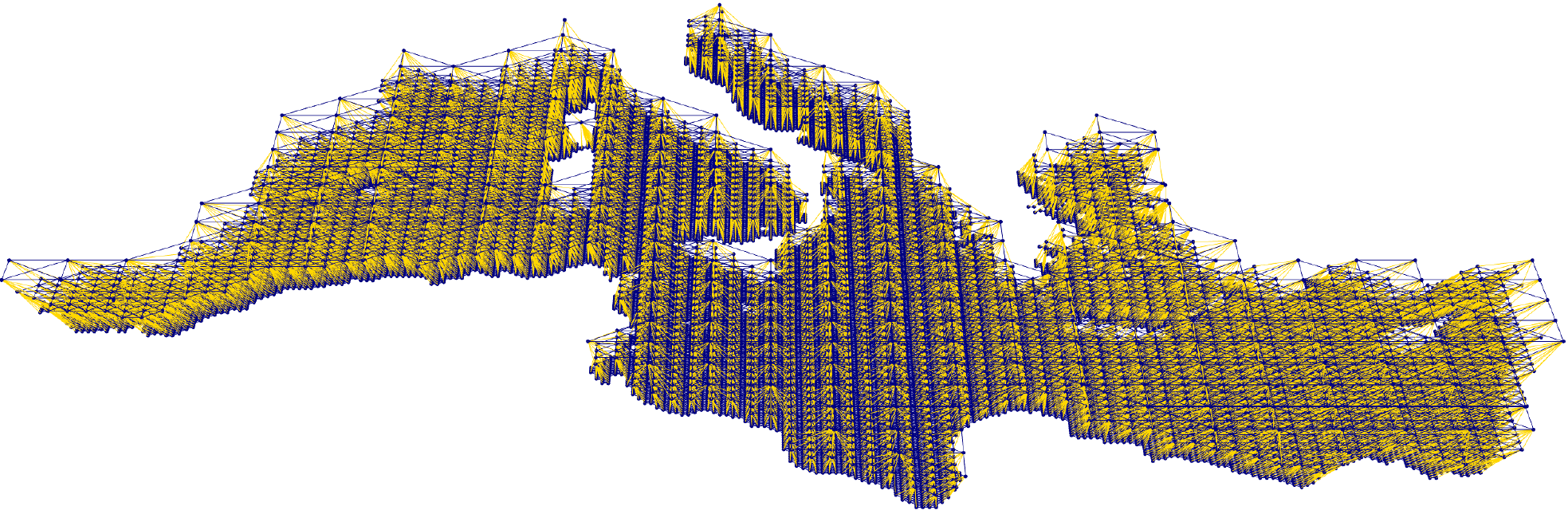}
    \caption{Ocean variables are encoded onto a hierarchical mesh of the Mediterranean Sea shown here. Each layer has a different resolution allowing for interactions at different scales between observables.}
    \label{fig:hi_graph}
\end{figure}

SeaCast operates on a multi-resolution graph representation of the Mediterranean Sea, coarser than the original grid, enabling the model to capture ocean dynamics across multiple spatial scales, while being computationally feasible to process on. Input variables are mapped onto a three-layer hierarchical graph, illustrated in Figure~\ref{fig:hi_graph}. Each graph layer captures ocean variability at a distinct resolution, allowing the model to learn both short and long range interactions that are useful for predicting future sea states.

Each node is connected via bidirectional edges to its immediate neighbors—horizontally, vertically, and diagonally. This connectivity pattern is repeated across three resolutions, with the spacing between nodes tripling at each successive level. Specifically, a node at resolution level $h$ is placed at the center of a $3 \times 3$ patch of nodes at resolution $h - 1$.

Table~\ref{tab:graph} summarizes the structure of the full graph used in SeaCast, including the number of nodes and edges at each layer, as well as the mapping components that connect the physical grid, mesh, and graph representations. The base layer, $\mathcal{G}_0$, corresponds to the finest resolution in the mesh. Two additional layers, $\mathcal{G}_1$ and $\mathcal{G}_2$, represent increasingly coarser resolutions, allowing for efficient message passing over larger spatial extents.

\begin{table}[h]
\centering
\caption{Number of nodes and edges in the sea graph.}
\begin{tabular}{@{}lll@{}}
\toprule
Graph                   & Nodes     & Edges \\
\midrule
$\mathcal{G}_0$         & 22677     & 174007 \\
$\mathcal{G}_{0, 1} / \mathcal{G}_{1,0}$ & -        & 22677  \\
$\mathcal{G}_1$         & 2515      & 18206    \\
$\mathcal{G}_{1, 2} / \mathcal{G}_{2, 1}$ & -        & 2515   \\
$\mathcal{G}_2$         & 272       & 1610 \\
\midrule
Mesh             & 25464    & 219015  \\
\midrule
$\mathcal{G}_{G2M}$     & -       & 542271 \\
$\mathcal{G}_{M2G}$     & -       & 579960 \\
\midrule
Grid                    & 144990   & - \\
\bottomrule
\end{tabular}
\label{tab:graph}
\end{table}

Inter-layer connections, such as $\mathcal{G}_{0,1}$ and $\mathcal{G}_{1,2}$, define bidirectional edges that facilitate hierarchical information flow between levels. Additionally, SeaCast includes mappings between the physical simulation grid and the mesh graph, denoted $\mathcal{G}_{G2M}$ (grid-to-mesh) and $\mathcal{G}_{M2G}$ (mesh-to-grid), which enable the model to interface with the ocean data grid.

\clearpage
\section{Evaluation metrics}

\paragraph{Root mean squared error (RMSE)} For each lead time $t$, variable $i$, and depth level $l$ the RMSE is defined as:
\begin{equation}
\text{RMSE}_{t, i, l}
=
\sqrt{
    \frac{1}{T_\text{eval}} \sum_{t_0=1}^{T_\text{eval}} \frac{1}{N_{l}} \sum_{v=1}^{N_{l}} a_v \left(\hat{X}_{v, i}^{t_0+t} - X_{v, i}^{t_0+t}\right)^2
}.
\end{equation}
Here, $t_0$ indexes forecast initialization times up to $T_\text{eval}$; $v$ indexes valid grid nodes at depth level $l$ up to $N_{l}$; $a_v$ is the cell‐area weight (normalized to unit mean); and $\hat X$ and $X$ denote the predicted and observed values, respectively.

\paragraph{Root mean squared error (RMSE), depth-averaged}  
For each lead time $t$ and variable $i$, the RMSE is defined as:
\begin{equation}
\text{RMSE}_{t, i}
=
\sqrt{
    \frac{1}{T_\text{eval}} \sum_{t_0=1}^{T_\text{eval}} \frac{1}{L_i} \sum_{l=1}^{L_i} \frac{1}{N_l} \sum_{v=1}^{N_l} a_v \left(\hat{X}_{v, i}^{t_0+t} - X_{v, i}^{t_0+t}\right)^2
}.
\end{equation}
Here, $l$ indexes the vertical levels associated with variable $i$ up to $L_i$, and the other variables are as defined above.

\paragraph{Root mean squared error (RMSE), per location}
We evaluate model outputs on observed SST also by breaking down RMSE per location. The spatial RMSE is calculated for lead time $t$, variable $i$, and grid point $v$ as:

\begin{equation}
\mathrm{RMSE}_{t,i,v}
=
\sqrt{
  \frac{1}{T_\text{eval}} \sum_{t_0=1}^{T_\text{eval}} \left(\hat{X}_{v,i}^{t_0+t} - X_{v,i}^{t_0+t}\right)^2
}.
\end{equation}

\paragraph{Heidke skill score (HSS)}

We evaluate the models' ability to detect temperature extremes for each lead time using The Heidke Skill Score (HSS) derived from the contingency table consisting of true positives (TP), false positives (FP), false negatives (FN), and true negatives (TN). HSS compares forecast accuracy to that expected by chance. It is defined by:
\begin{equation}
    \text{HSS} = \frac{2\,(TP \cdot TN - FP \cdot FN)}{(TP+FN)(FN+TN) + (TP+FP)(FP+TN)}.
\end{equation}
An HSS of 1 indicates a perfect forecast, 0 indicates no better than chance, and negative values indicate performance worse than a random forecast.

\paragraph{Anomaly Correlation Coefficient (ACC), depth-averaged}
For each lead time $t$ and multi-level variable $i$, we define the forecast and analysis anomalies as:
\begin{equation}
A_{v,i}^{t_0+t} = \hat{X}_{v,i}^{t_0+t} - C_{v,i}^{t_0+t}, \quad
B_{v,i}^{t_0+t} = X_{v,i}^{t_0+t} - C_{v,i}^{t_0+t}.
\end{equation}
The anomaly correlation coefficient is then computed as:
\begin{equation}
\text{ACC}_{t,i}
=
\frac{1}{T_\text{eval}} \sum_{t_0=1}^{T_\text{eval}}
\frac{
  \frac{1}{L_i}\sum_{l=1}^{L_i} \sum_{v=1}^{N_l} a_v
  A_{v,i}^{t_0+t} B_{v,i}^{t_0+t}
}{
  \sqrt{\left[
    \frac{1}{L_i}\sum_{l=1}^{L_i} \sum_{v=1}^{N_l} a_v
    \left( A_{v,i}^{t_0+t} \right)^2
  \right]
  \left[
    \frac{1}{L_i}\sum_{l=1}^{L_i} \sum_{v=1}^{N_l} a_v
    \left( B_{v,i}^{t_0+t} \right)^2
  \right]}
}.
\end{equation}
Here, $C_{v,i}^{t_0+t}$ is the daily climatology at grid point $v$ and time $t_0 + t$, and the other variables are as defined above. ACC hence measures
the correlation between the deviation of the prediction and ground truth from daily climatology.

\clearpage
\section{Additional results}

\subsection{In-situ evaluation}

To assess forecasted depth-wise variables on ground-truth beyond gridded analysis data, we compare them against quality controlled in-situ measurements from the Mediterranean Sea. The in-situ data is sourced from the CMEMS In Situ Thematic Assembly Centre, which compiles observations from a broad network of national and international monitoring systems. To ensure consistency with the model output, the observations are spatially and temporally bin-averaged, following the SeaCast simulation grid, to produce daily mean values. Forecast fields are interpolated bilinearly to the latitude–longitude coordinates of the observations and linearly in the vertical to match the measured depths.

\begin{figure}[h]
\centering
\includegraphics[width=.8\textwidth]{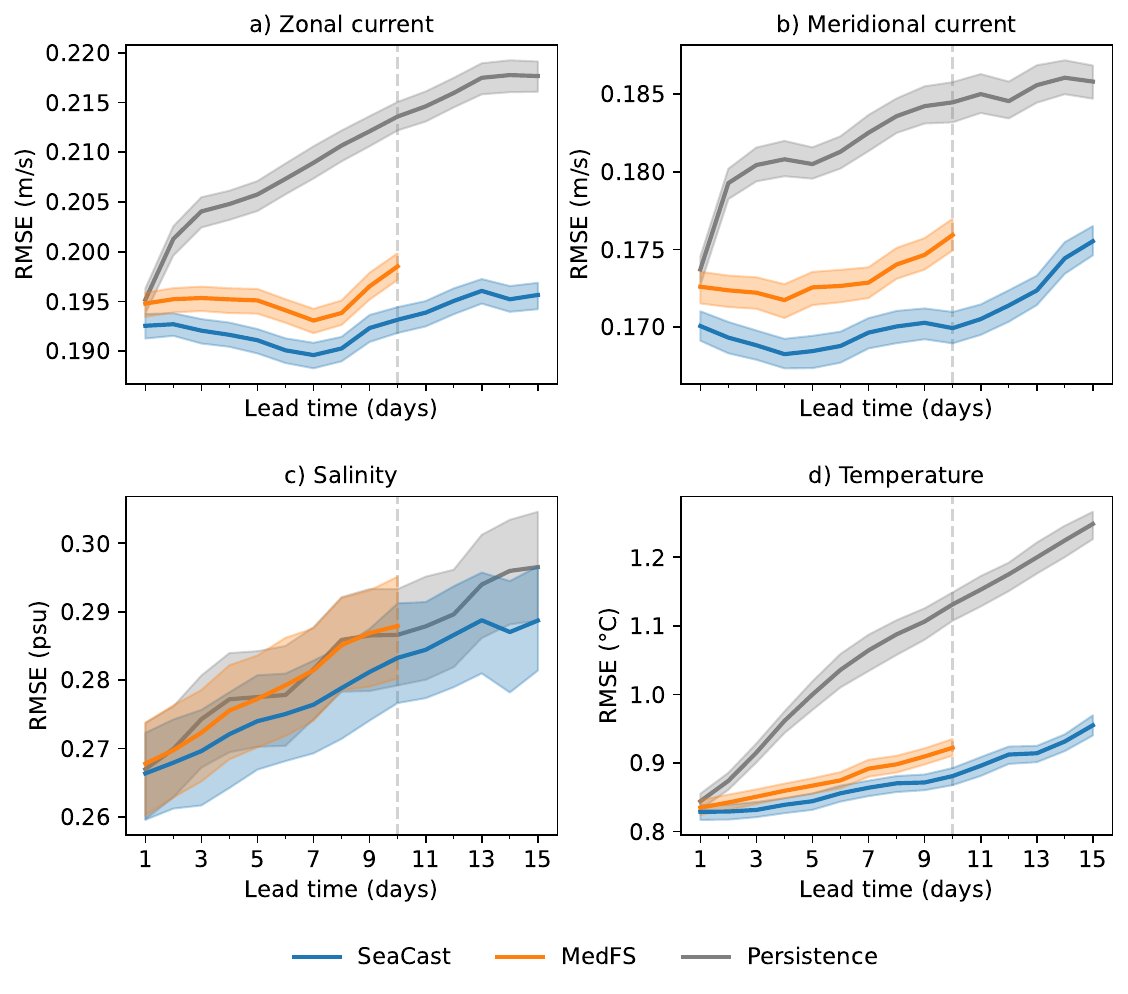}
\caption{RMSE as a function of lead time for SeaCast forecasts compared to in-situ measurements. Results are shown for zonal and meridional currents, salinity, and temperature. The shading corresponds to the $50\%$ confidence intervals estimated via bootstrapping.}
\label{fig:in_situ}
\end{figure}

\clearpage

\subsection{Analysis evaluation}

Here, model forecasts are evaluated against analysis fields, using both depth-averaged ACC and RMSE at every other forecasted depth level. To account for the impact of the weekly data assimilation cycle, we also compute errors separately for forecasts initialized on the best performing day of the week, which is on Tuesday. Given the sizeable improvements we see below, these results make the case for performing assimilation more often, or at least aim to do so in the future. Zonal and meridional currents behave so similarly that only the former is included for the RMSE evaluation.

\begin{figure}[ht]
\centering
\includegraphics[width=.8\textwidth]{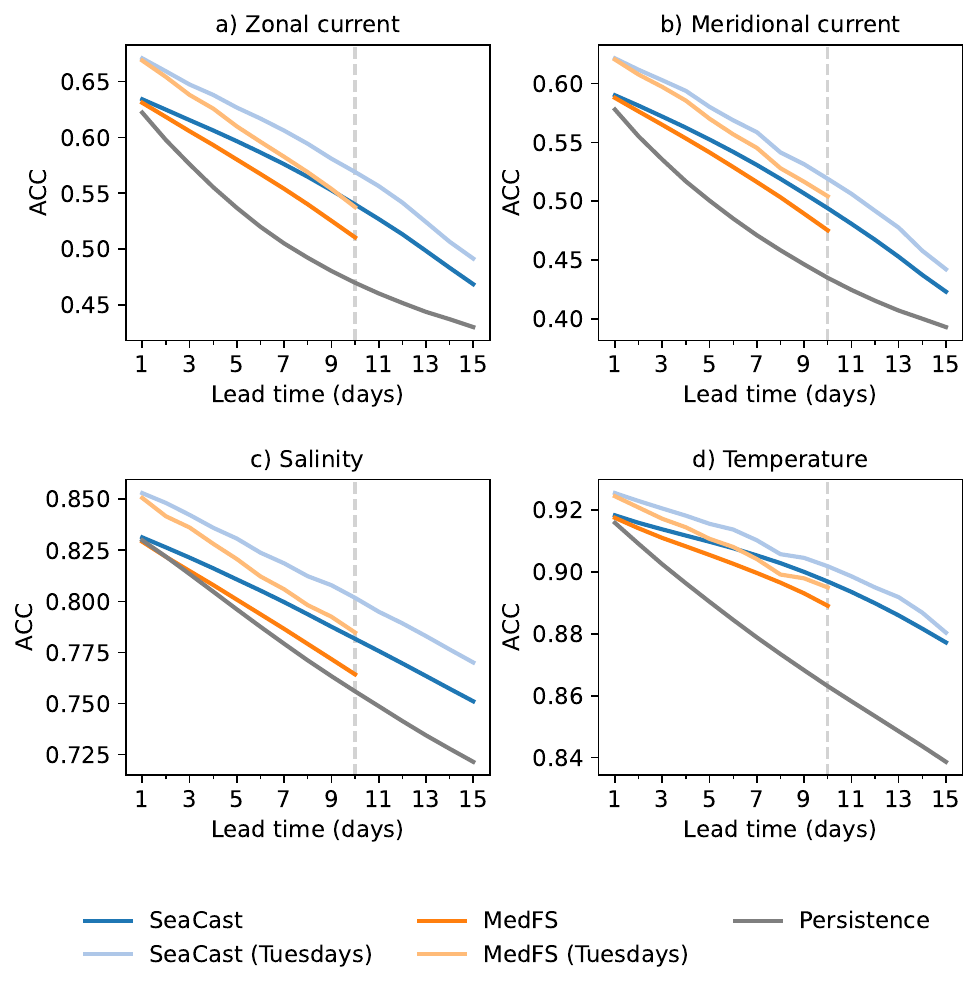}
\caption{ACC versus lead time for SeaCast, MedFS, and a persistence baseline. Additional SeaCast and MedFS results from Tuesday initializations are included to account for peak performance due to weekly data assimilation.}
\label{fig:acc_avg}
\end{figure}

\begin{figure}[ht]
\centering
\includegraphics[width=\textwidth]{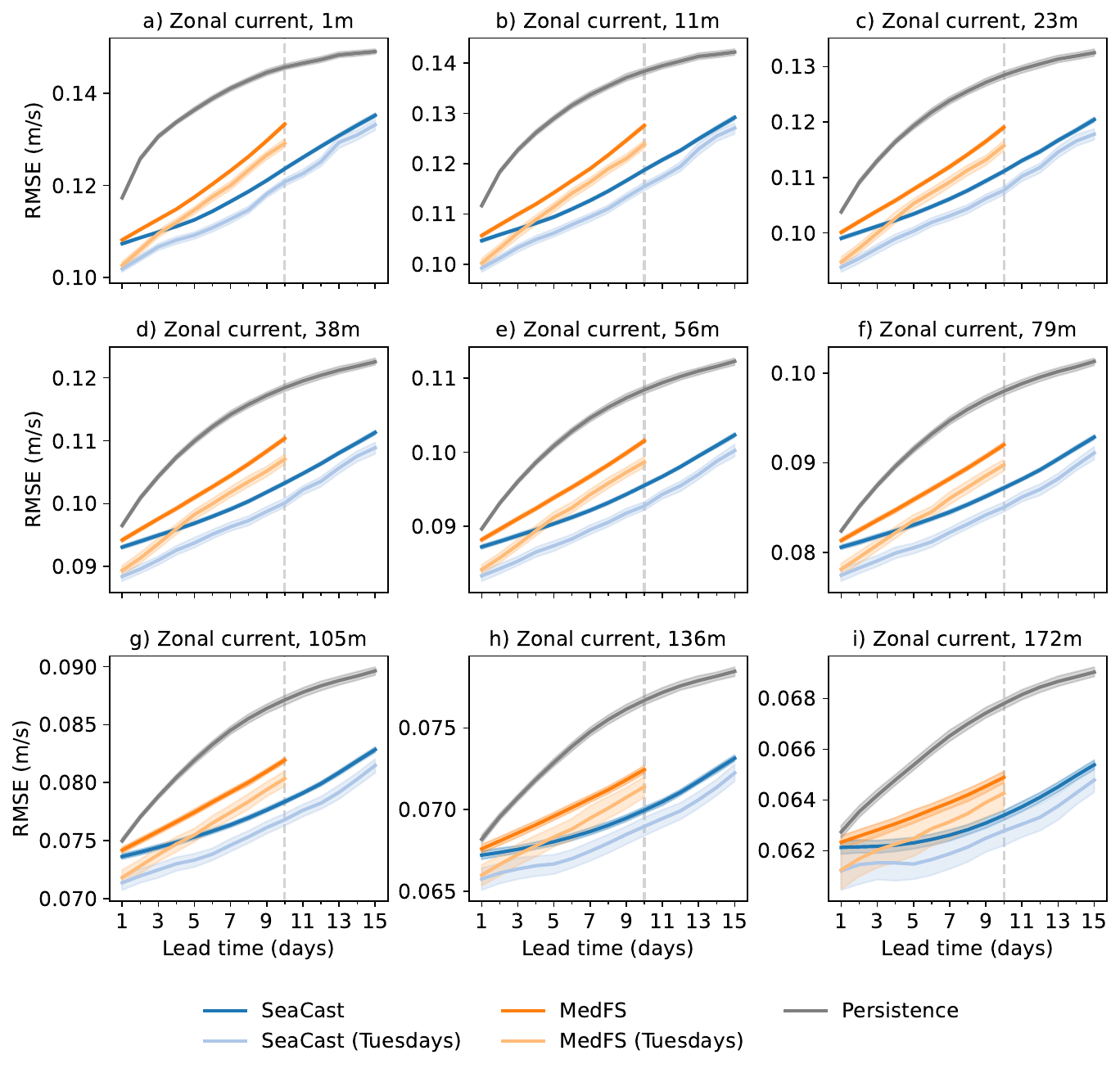}
\caption{RMSE of zonal currents versus lead time across depth levels. Each subplot corresponds to a different depth. Comparisons include SeaCast, MedFS, and persistence. Additional SeaCast and MedFS results from Tuesday initializations are included to account for peak performance due to weekly data assimilation. The shading corresponds to the $50\%$ confidence intervals estimated via bootstrapping.}
\label{fig:uo_rmse_models}
\end{figure}

\begin{figure}[ht]
\centering
\includegraphics[width=\textwidth]{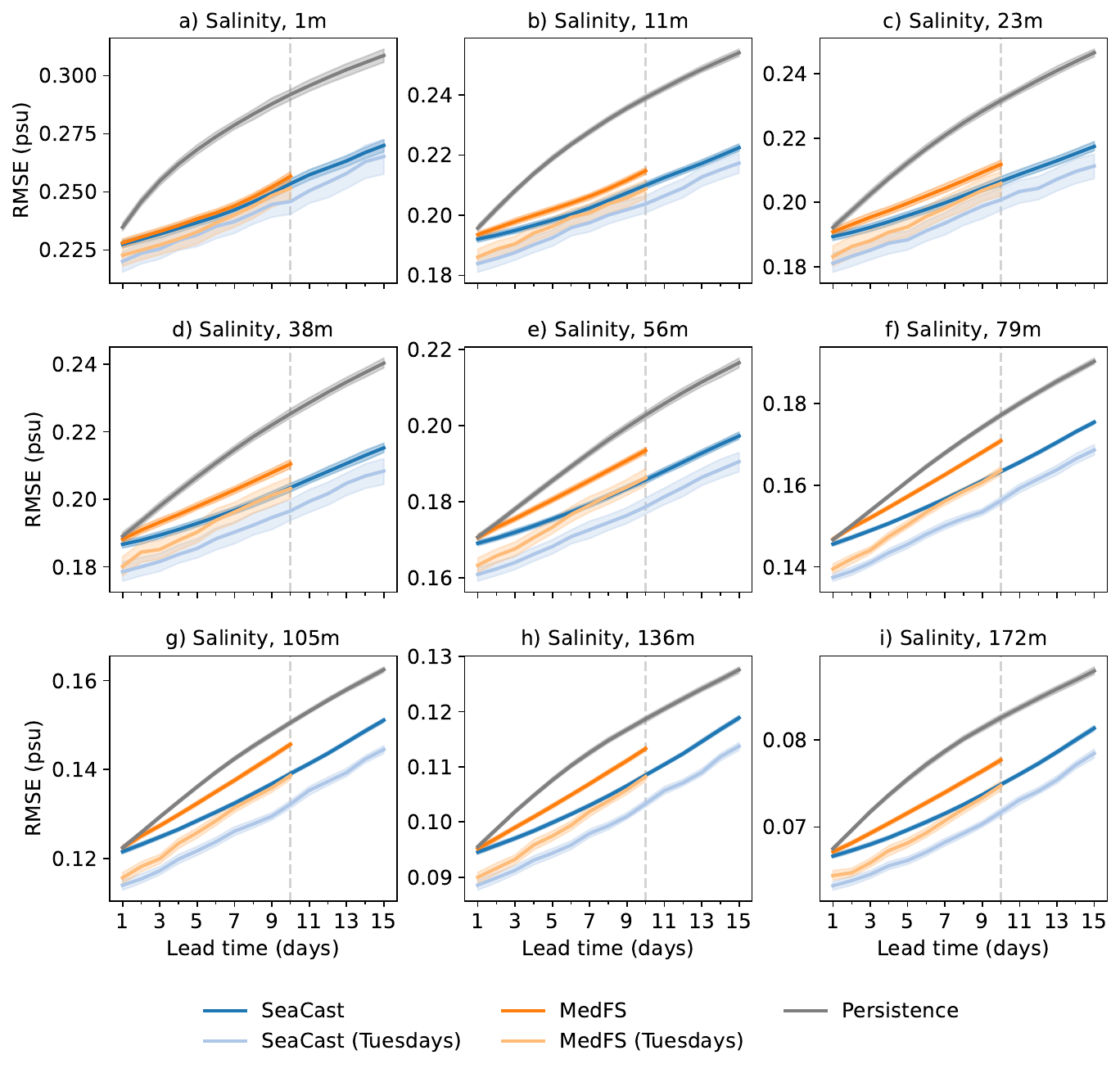}
\caption{RMSE of salinity versus lead time across depth levels. Each subplot represents a different depth, comparing SeaCast, MedFS, and persistence. Additional SeaCast and MedFS results from Tuesday initializations are included to account for peak performance due to weekly data assimilation. The shading corresponds to the $50\%$ confidence intervals estimated via bootstrapping.}
\label{fig:so_rmse_models}
\end{figure}

\begin{figure}[ht]
\centering
\includegraphics[width=\textwidth]{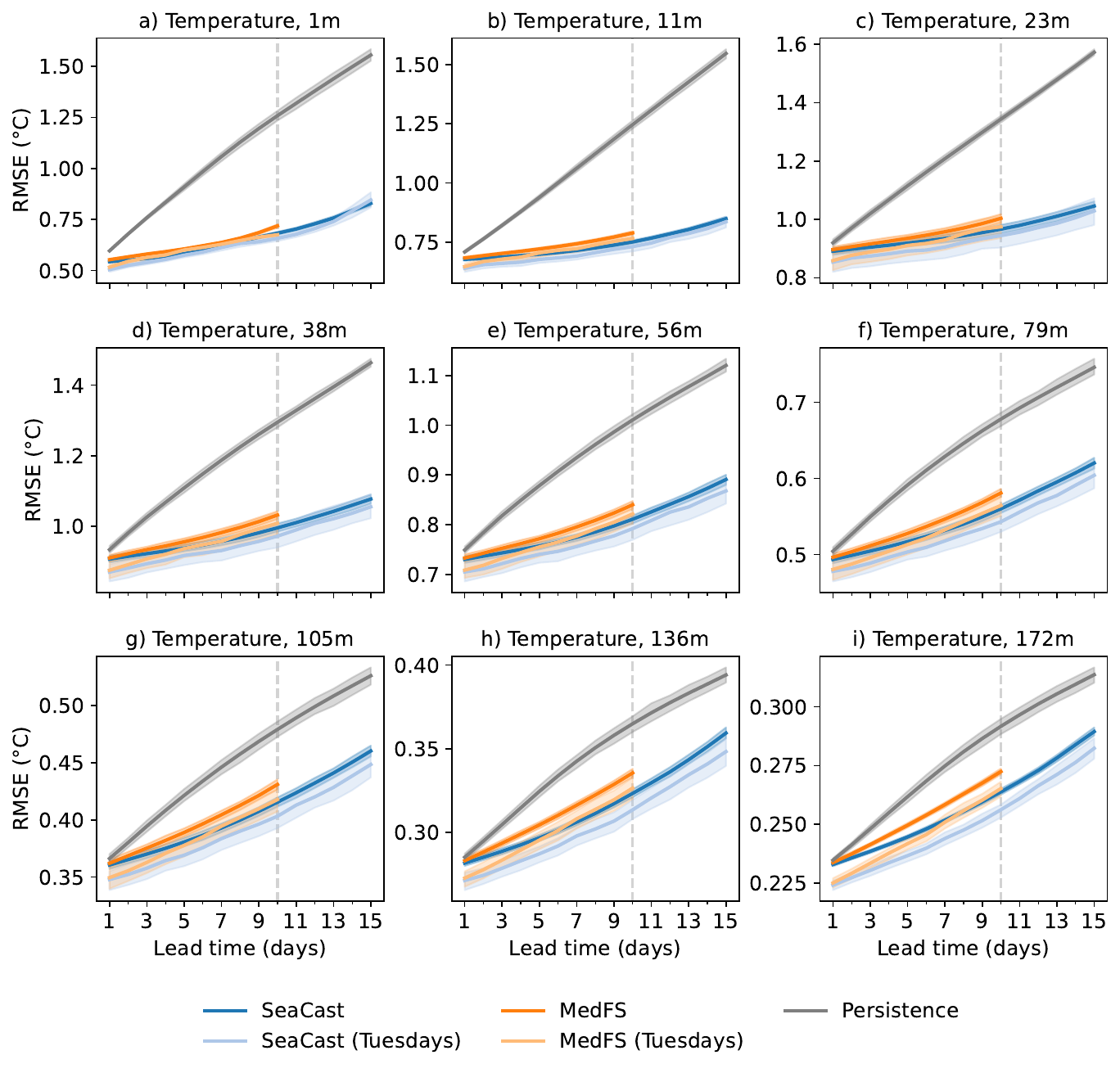}
\caption{RMSE of temperature versus lead time across depth levels. Each subplot corresponds to a distinct depth, showing performance of SeaCast, MedFS, and persistence. Additional SeaCast and MedFS results from Tuesday initializations are included to account for peak performance due to weekly data assimilation. The shading corresponds to the $50\%$ confidence intervals estimated via bootstrapping.}
\label{fig:thetao_rmse_models}
\end{figure}

\clearpage

\subsection{Vertical error profiles}

To better understand how forecast skill varies with depth, we analyze RMSE profiles across the vertical column for key ocean variables. These profiles are averaged over lead times only to reveal depth-dependent behavior. SeaCast initialized from the best available analysis fields serves as a lower bound on forecast error, while a persistence baseline provides an upper bound. Zonal and meridional currents behave so similarly that only the former is included here.

\begin{figure}[ht]
\centering
\includegraphics[width=\textwidth]{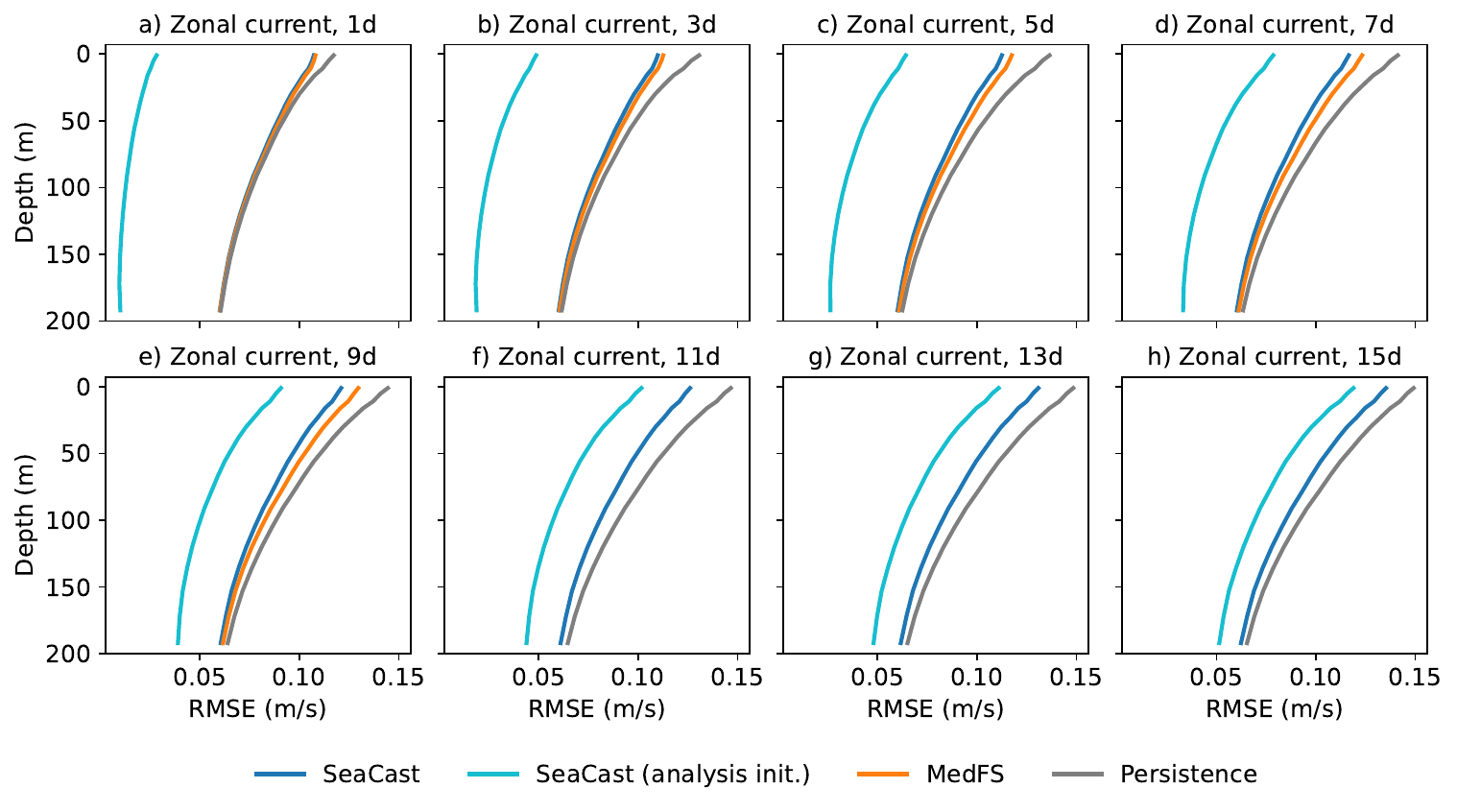}
\caption{Vertical profile of RMSE for zonal currents averaged across lead times. SeaCast initialized from analysis fields and a persistence model are included to show upper and lower performance bounds across depths.}
\label{fig:uo_rmse_depth}
\end{figure}

\begin{figure}[ht]
\centering
\includegraphics[width=\textwidth]{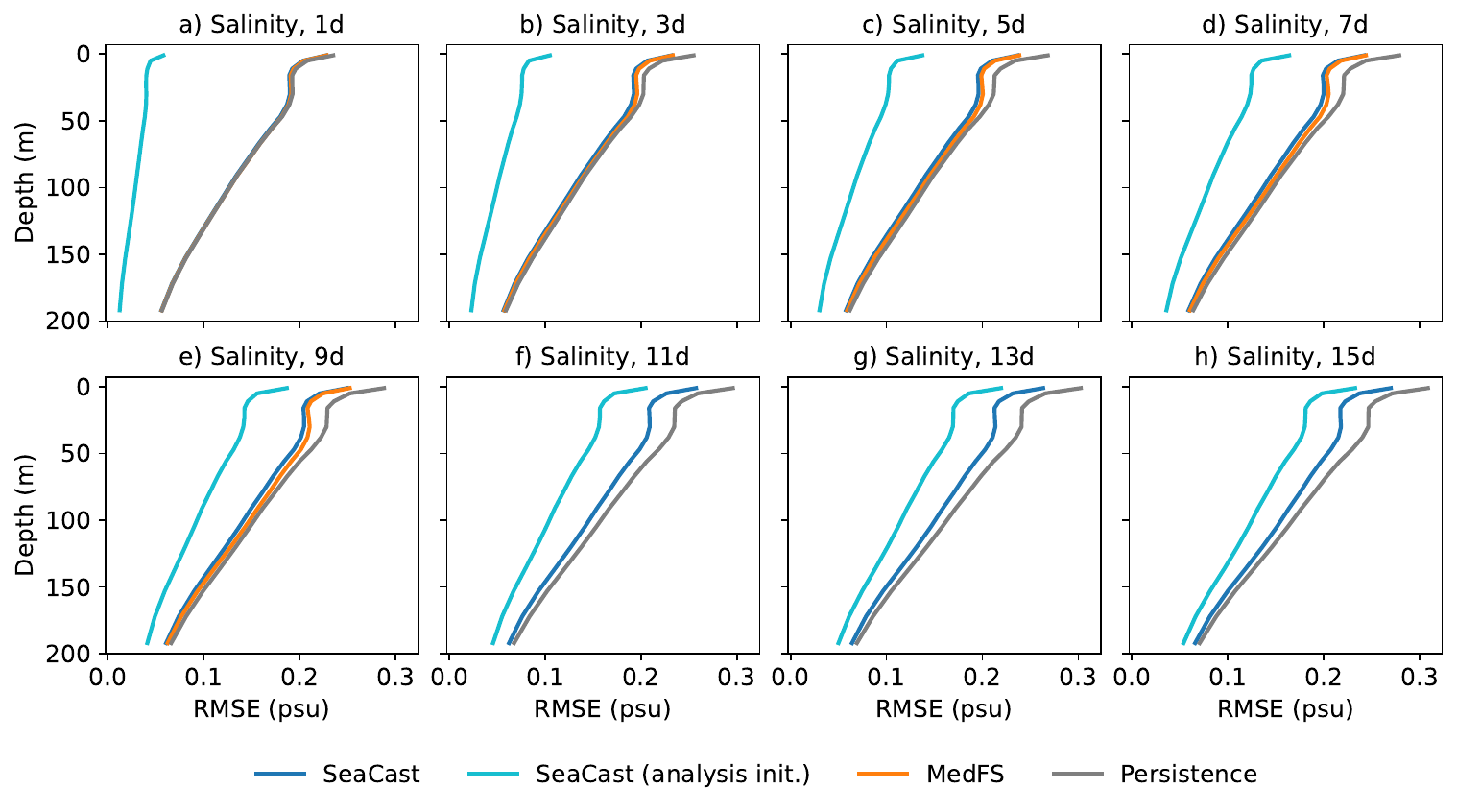}
\caption{Vertical profile of RMSE for salinity, averaged over all lead times. SeaCast initialized from analysis fields and a persistence model are included to show upper and lower performance bounds across depths.}
\label{fig:so_rmse_depth}
\end{figure}

\begin{figure}[ht]
\centering
\includegraphics[width=\textwidth]{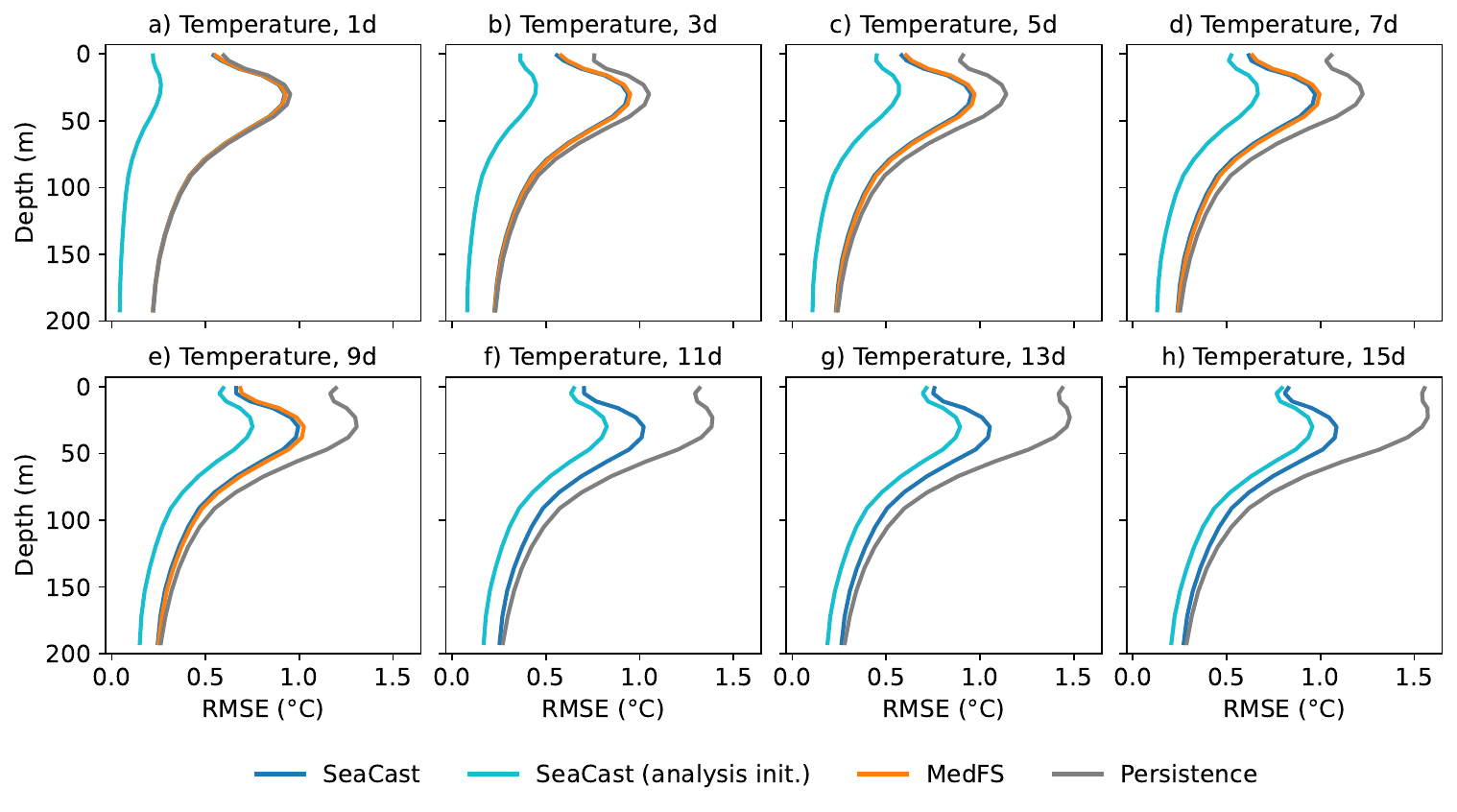}
\caption{Vertical profile of RMSE for temperature, averaged across lead times. SeaCast initialized from analysis fields and a persistence model are included to show upper and lower performance bounds across depths.}
\label{fig:thetao_rmse_depth}
\end{figure}

As seen in Figures~\ref{fig:so_rmse_depth} and~\ref{fig:thetao_rmse_depth}, temperature, and salinity to a small extent, exhibit a bump in RMSE between approximately 10 and 60 meters depth. This depth range often corresponds to the base of the mixed layer during late summer-autumn seasons (which comprise a majority of the evaluation set), a zone prone to rapid transitions in stratification driven by atmospheric forcing or mesoscale variability, dynamics that remain challenging to capture accurately in simulations. Figure~\ref{fig:uo_rmse_depth} shows that RMSE for zonal currents does not exhibit a similar bump, indicating that errors in current forecasts are more vertically uniform and less sensitive to stratification-related biases compared to scalar fields like temperature and salinity.

\clearpage

\subsection{Effect of atmospheric forcing}

To examine the sensitivity of SeaCast to surface atmospheric inputs, we perform a series of ablation studies in which individual atmospheric fields are permuted during inference. The normalized RMSE differences reveal the relative importance of each forcing variable and their influence on forecast accuracy at different depths and lead times. Zonal and meridional currents behave so similarly that only the former is included here. We can see a clear trend of the atmospheric forcing being more impactful closer to the sea surface in the plots below.

\begin{figure}[ht]
\centering
\includegraphics[width=\textwidth]{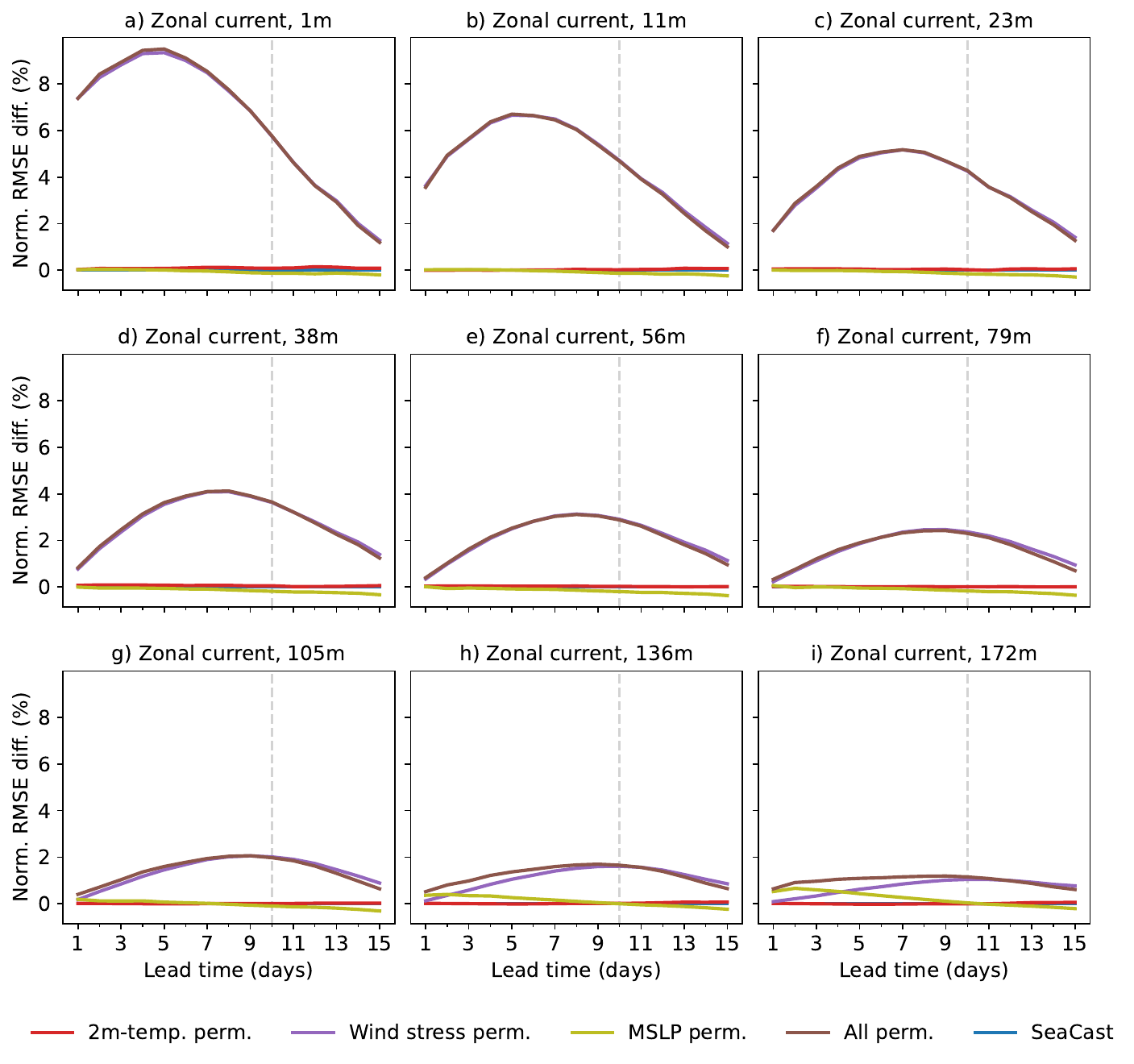}
\caption{Normalized RMSE difference for zonal currents per lead time across depths. Results show the impact of permuting individual atmospheric forcing fields relative to the original SeaCast configuration.}
\label{fig:uo_norm_rmse_diff_forcing}
\end{figure}

\begin{figure}[ht]
\centering
\includegraphics[width=\textwidth]{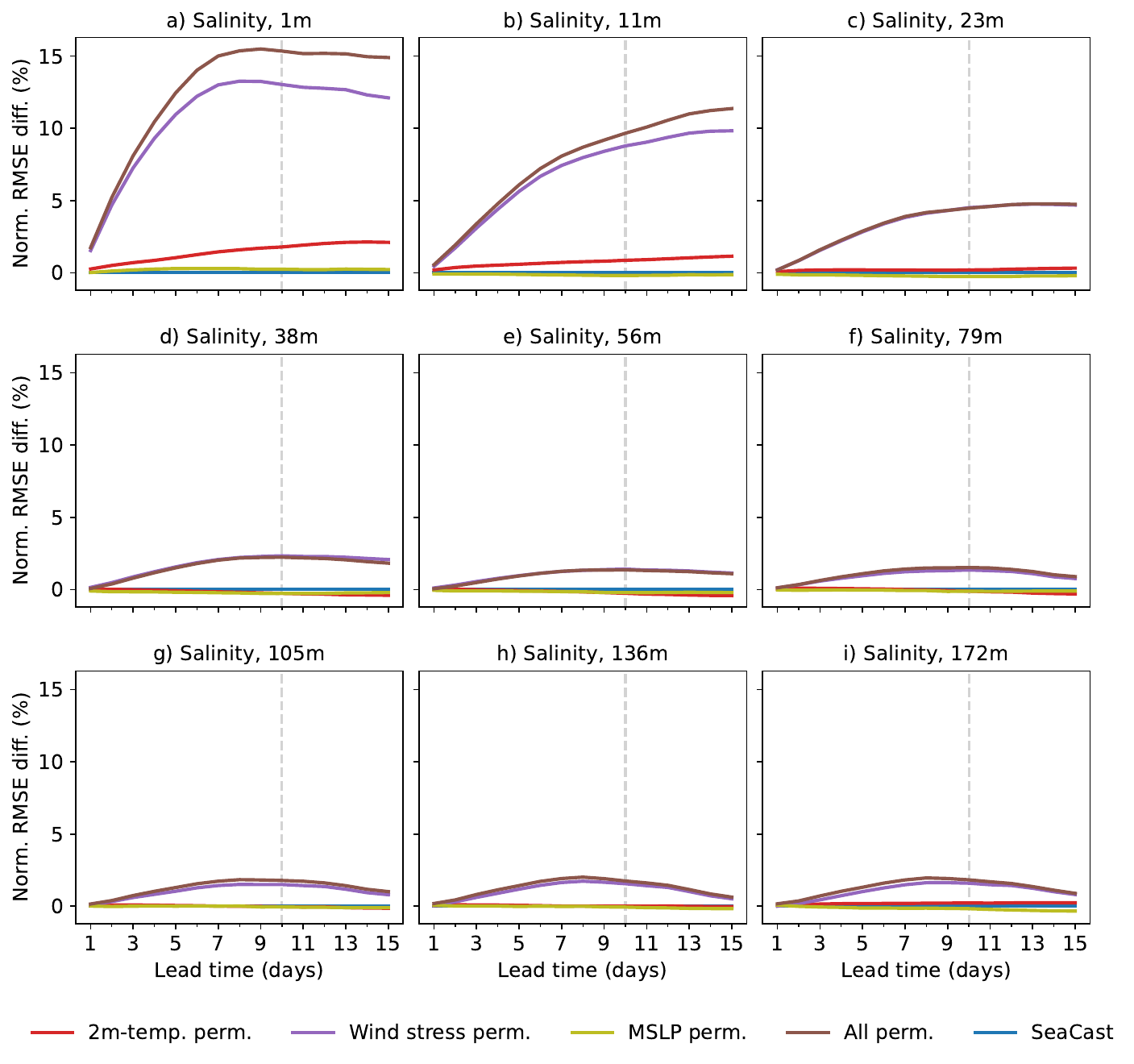}
\caption{Normalized RMSE difference for salinity per lead time across depths. Comparisons show the impact of permuted atmospheric forcings versus the unperturbed SeaCast baseline.}
\label{fig:so_norm_rmse_diff_forcing}
\end{figure}

\begin{figure}[ht]
\centering
\includegraphics[width=\textwidth]{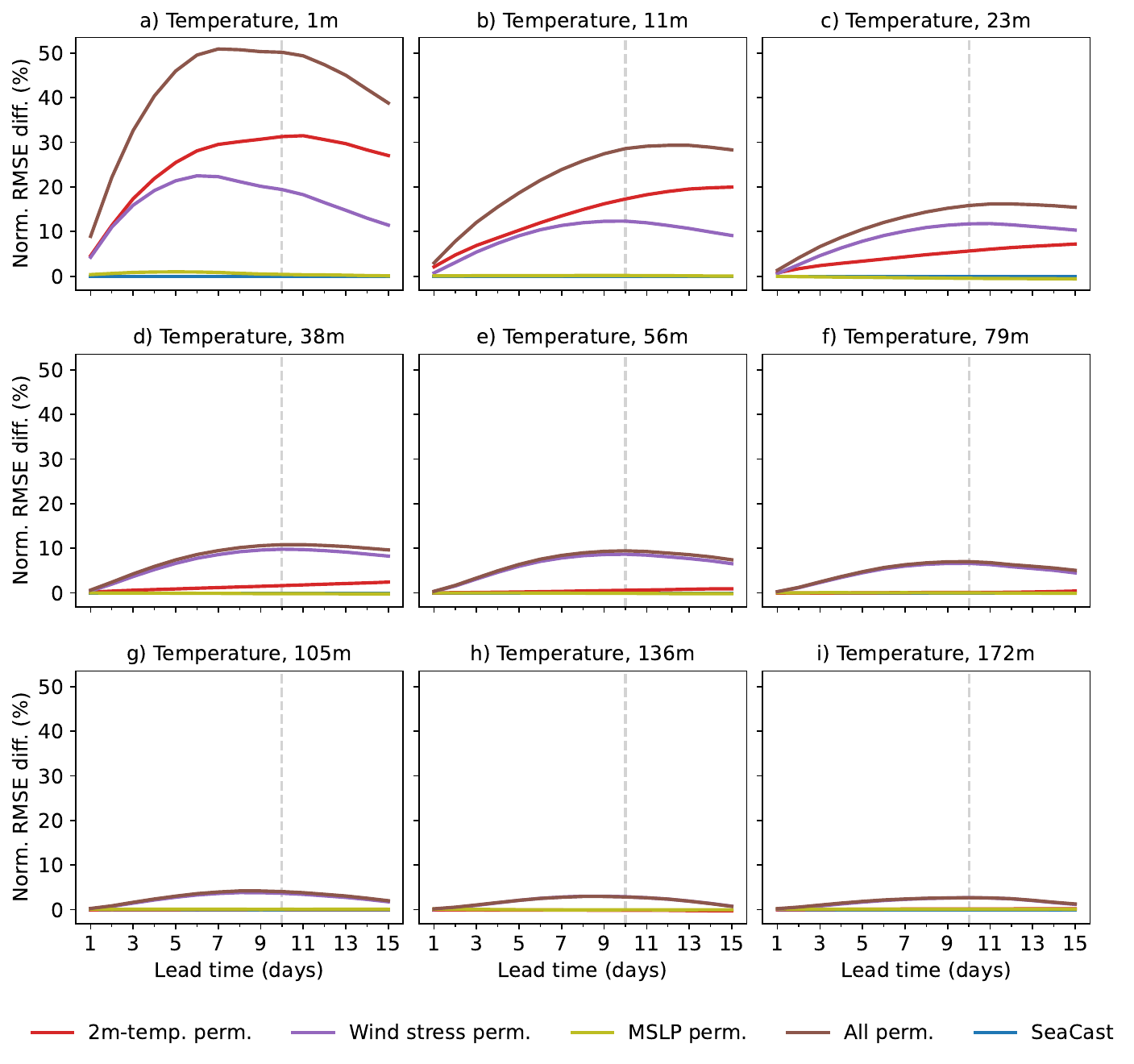}
\caption{Normalized RMSE difference for temperature per lead time across depths. Each line shows the effect of permuting one atmospheric variable compared to the original SeaCast model.}
\label{fig:thetao_norm_rmse_diff_forcing}
\end{figure}

\clearpage

\subsection{Effect of training period}

Here, we explore how extending the training period affects forecast performance by comparing SeaCast models trained on 35 years and 8 years of reanalysis data (w/o finetuning) to models that are further finetuned on 2 years of analysis data. The forecasts are benchmarked against the persistence model using the normalized RMSE difference as skill metric for each lead time. Zonal and meridional currents behave so similarly that only the former is included here.

\begin{figure}[ht]
\centering
\includegraphics[width=\textwidth]{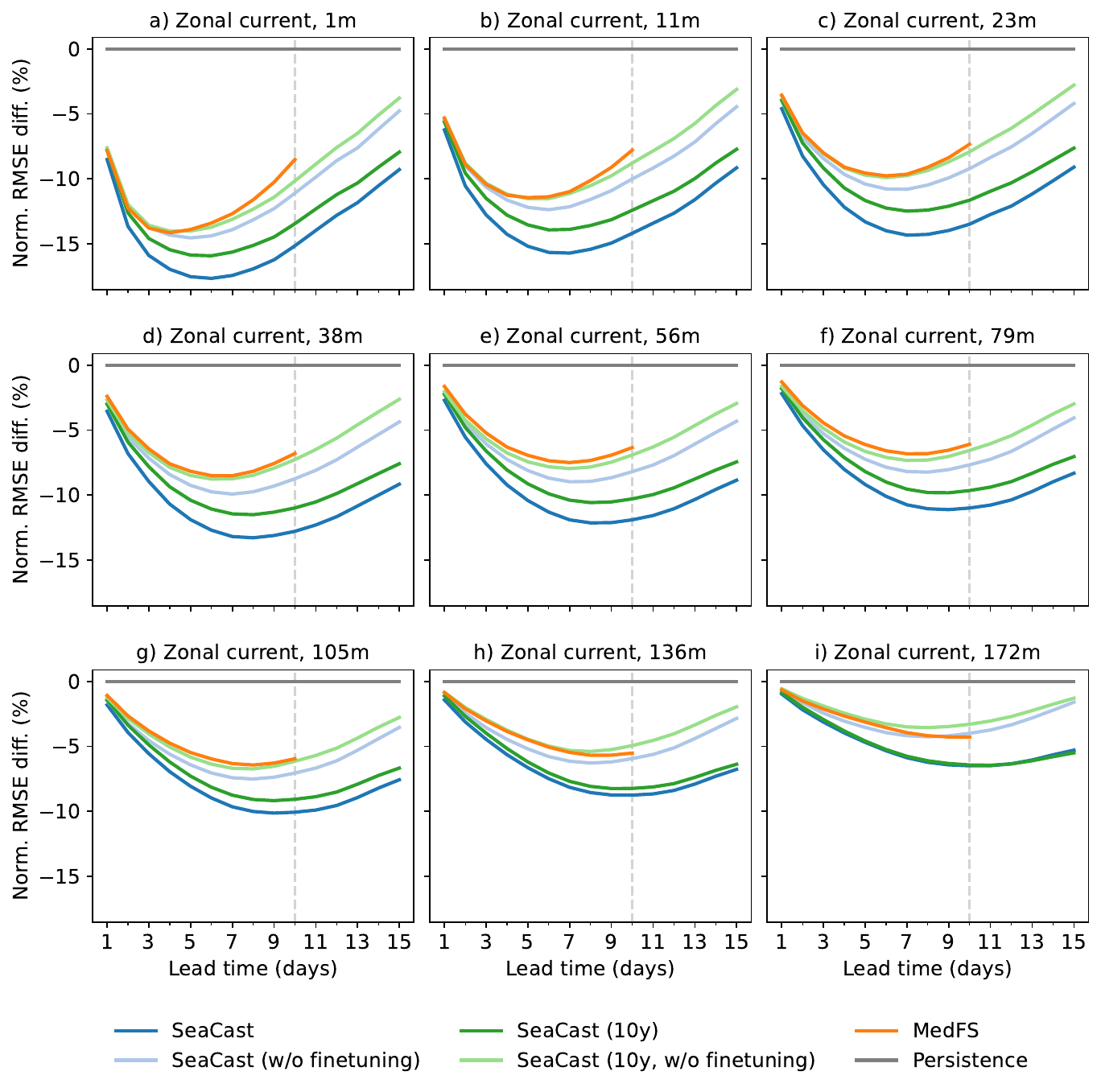}
\caption{Normalized RMSE difference for zonal current as a function of lead time, comparing SeaCast models trained on different time spans against the persistence baseline.}
\label{fig:uo_norm_rmse_diff_persistence}
\end{figure}

\begin{figure}[ht]
\centering
\includegraphics[width=\textwidth]{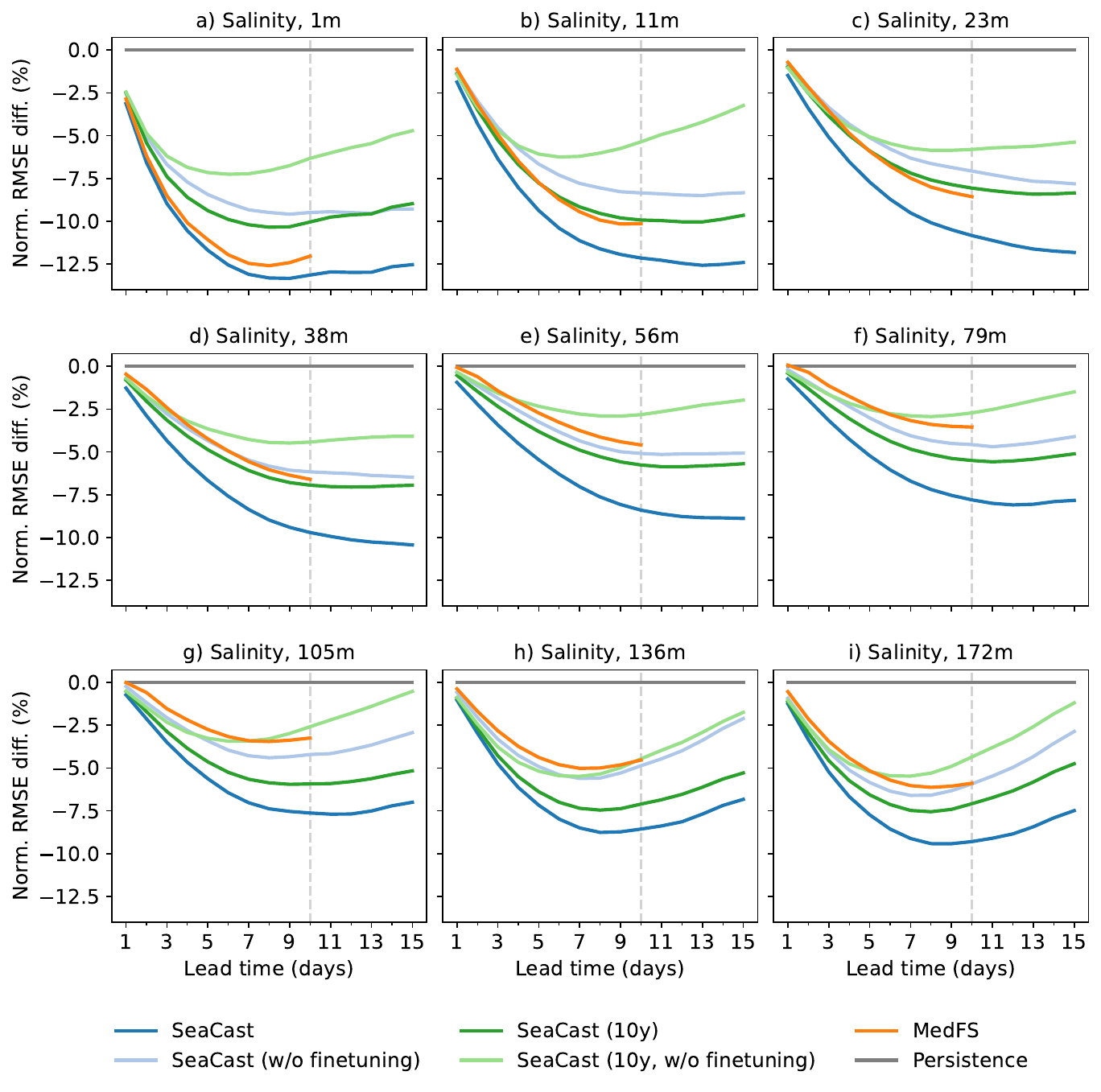}
\caption{Normalized RMSE difference for salinity as a function of lead time, comparing SeaCast models trained on different time spans against the persistence baseline.}
\label{fig:so_norm_rmse_diff_persistence}
\end{figure}

\begin{figure}[ht]
\centering
\includegraphics[width=\textwidth]{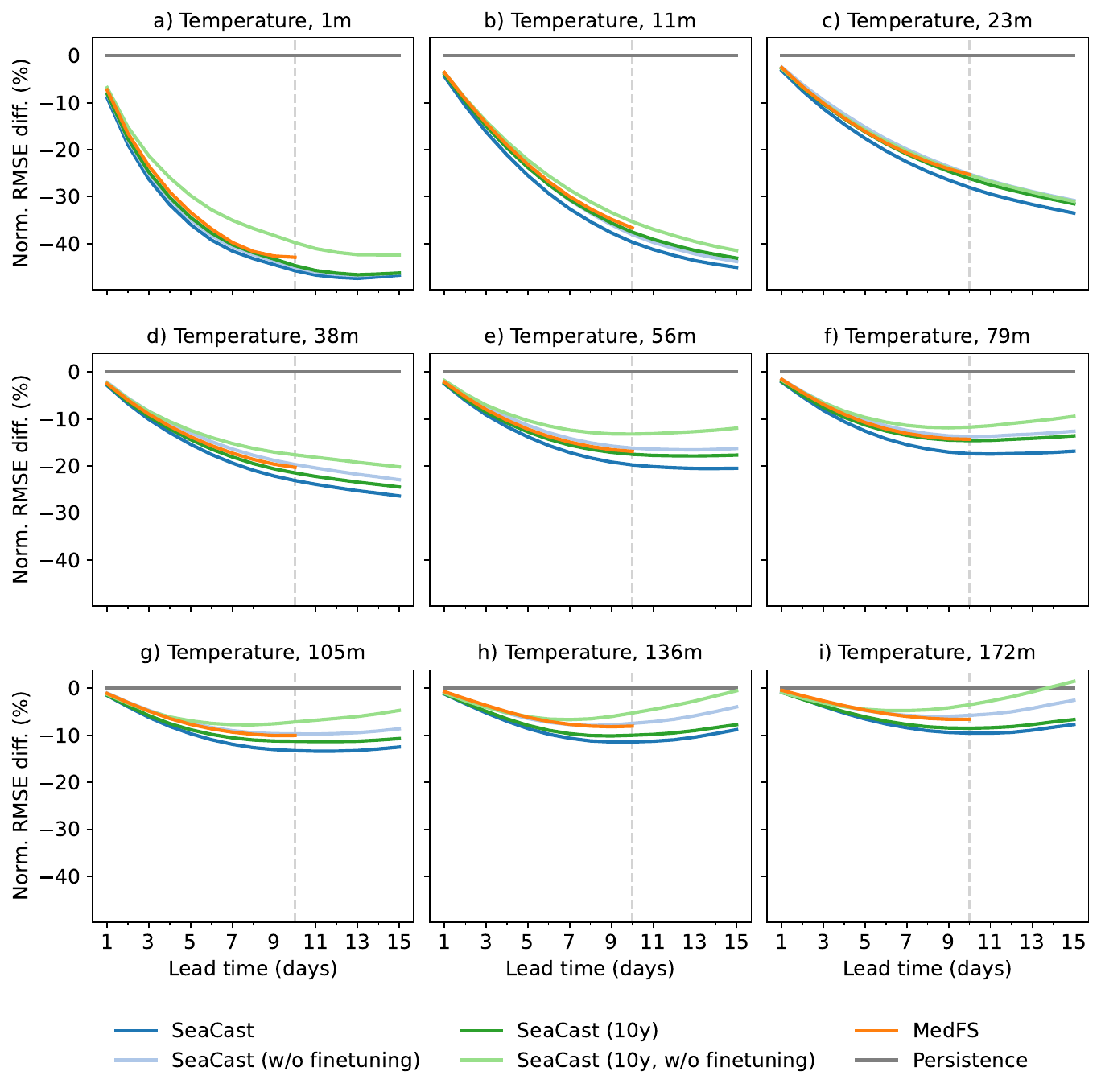}
\caption{Normalized RMSE difference for temperature as a function of lead time, comparing SeaCast models trained on different time spans against the persistence baseline.}
\label{fig:thetao_norm_rmse_diff_persistence}
\end{figure}

\clearpage

\section{Example forecasts}

To qualitatively illustrate SeaCast’s forecasting capabilities, we show example outputs for a selection of fields. Forecasts are initialized from operational simulation states on October 1st, 2024, and include horizontal maps of currents, temperature, salinity at 11\,m depth, as well as SSH. The bias is calculated towards the corresponding analysis fields.

\begin{figure}[ht]
\centering
\includegraphics[width=\textwidth]{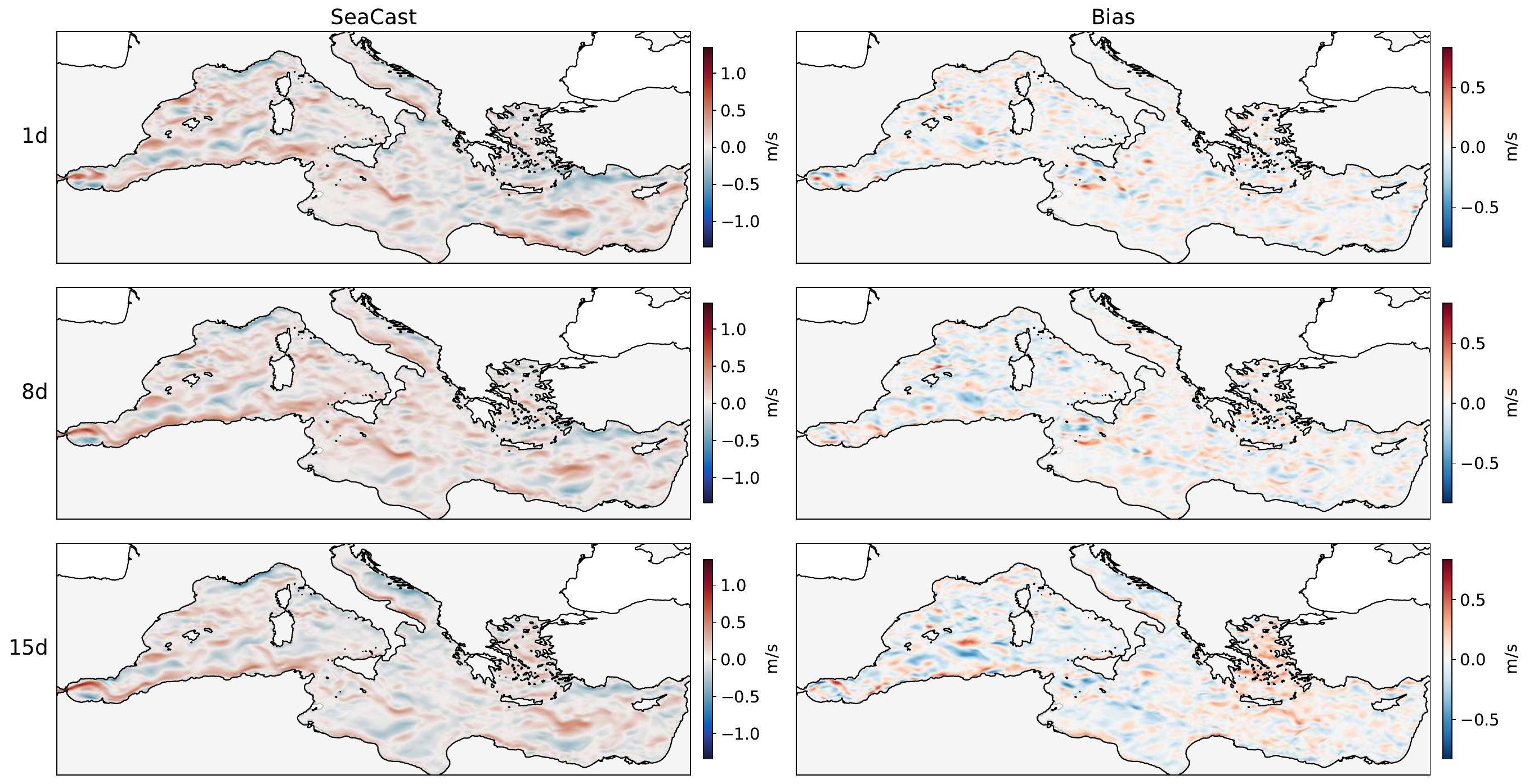}
\caption{SeaCast forecast of zonal current at 11\,m depth, initialized on October 1st, 2024.}
\label{fig:seacast_uo}
\end{figure}

\begin{figure}[ht]
\centering
\includegraphics[width=\textwidth]{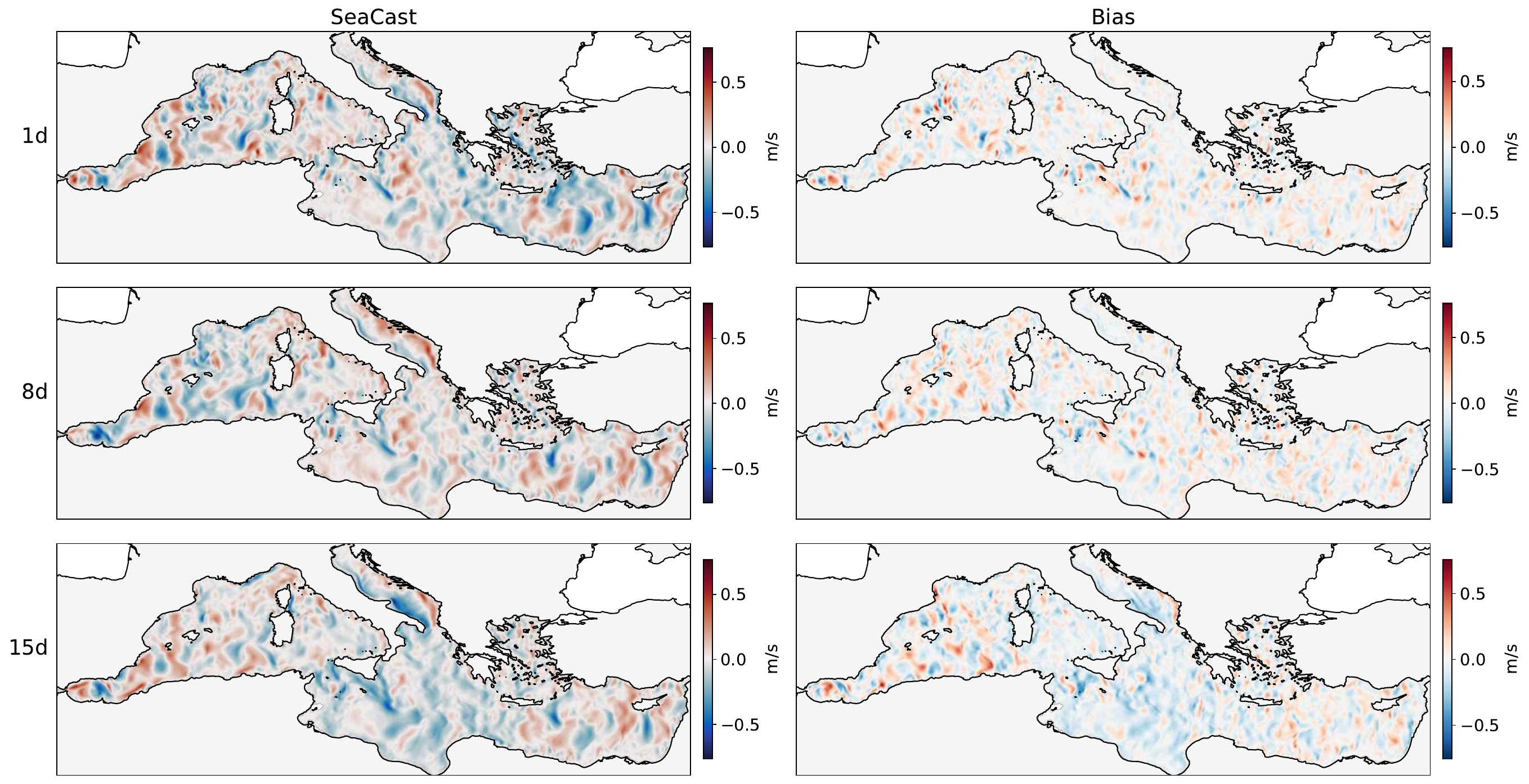}
\caption{SeaCast forecast of meridional current at 11\,m depth, initialized on October 1st, 2024.}
\label{fig:seacast_vo}
\end{figure}

\begin{figure}[ht]
\centering
\includegraphics[width=\textwidth]{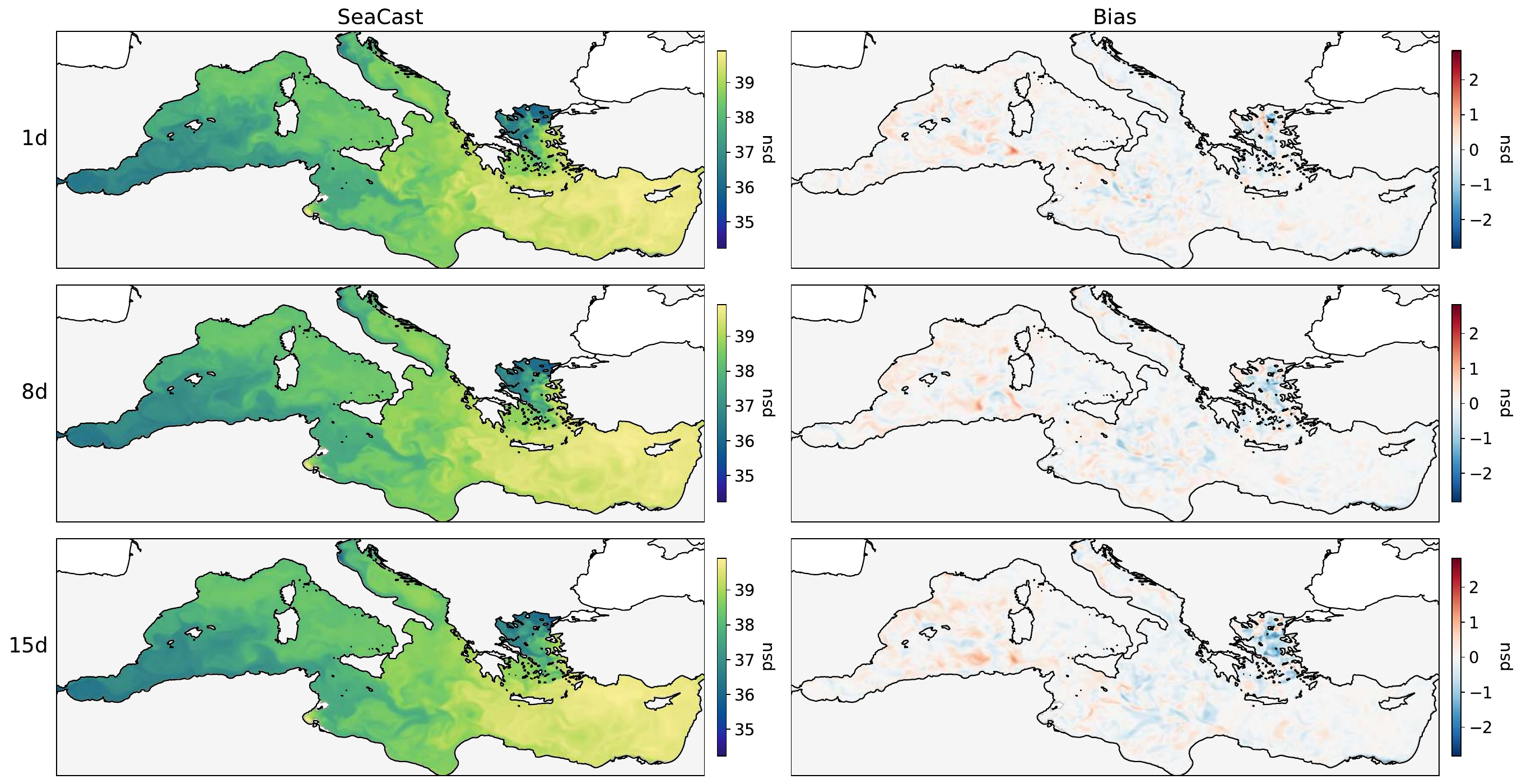}
\caption{SeaCast forecast of salinity at 11\,m depth, initialized on October 1st, 2024.}
\label{fig:seacast_so}
\end{figure}

\begin{figure}[ht]
\centering
\includegraphics[width=\textwidth]{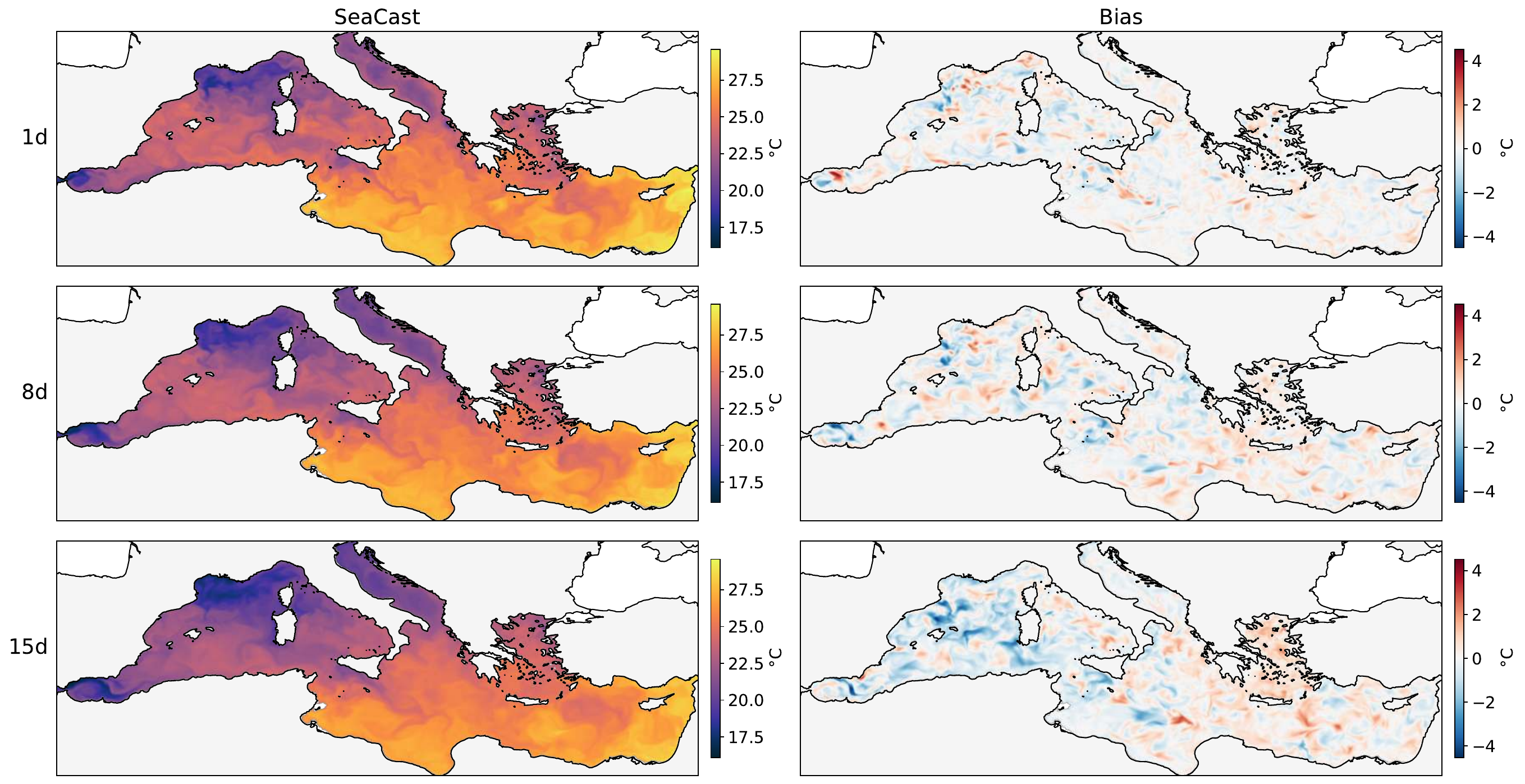}
\caption{SeaCast forecast of temperature at 11\,m depth, initialized on October 1st, 2024.}
\label{fig:seacast_thetao}
\end{figure}

\begin{figure}[ht]
\centering
\includegraphics[width=\textwidth]{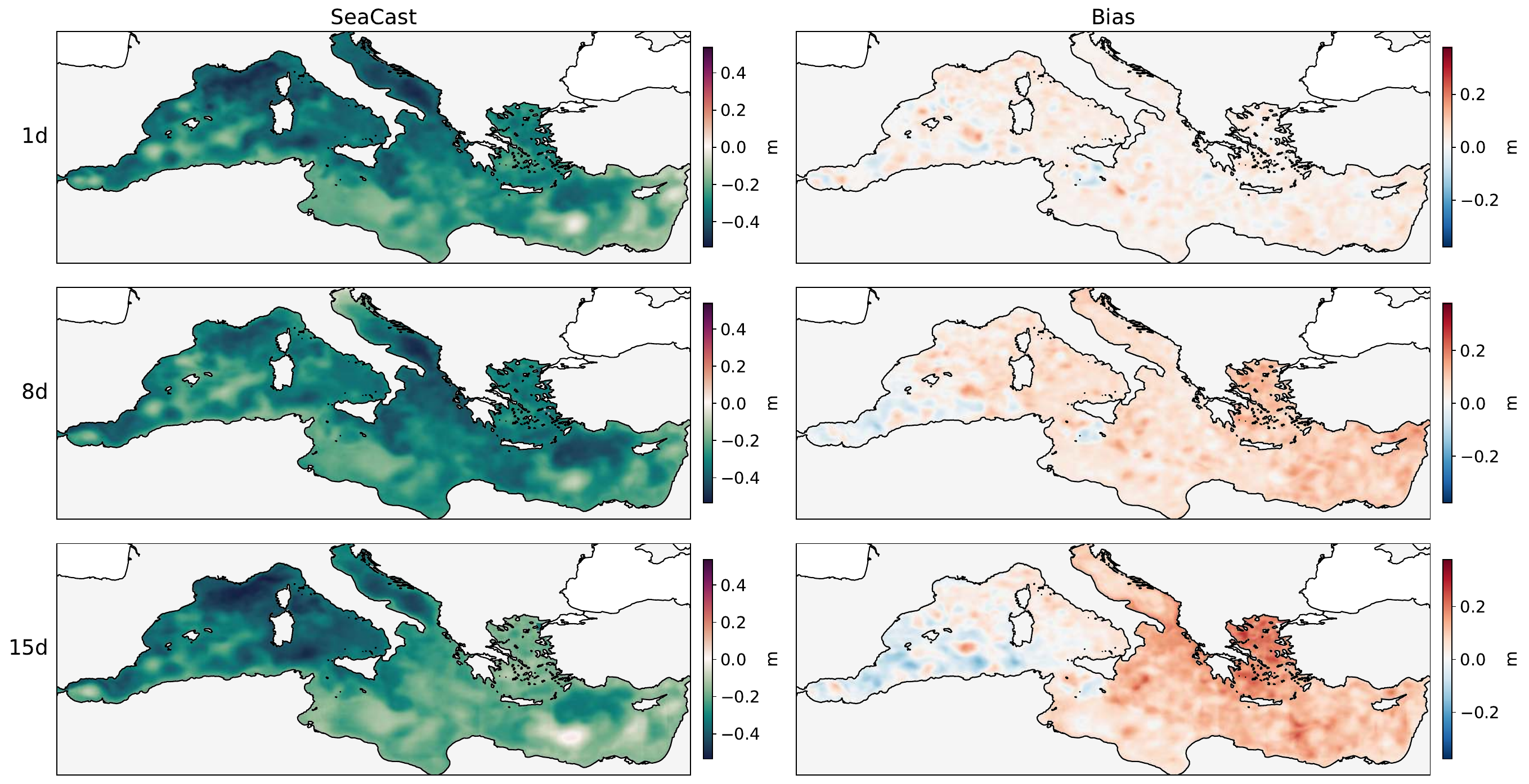}
\caption{SeaCast forecast of SSH, initialized on October 1st, 2024.}
\label{fig:seacast_zos}
\end{figure}